\documentclass[11pt, a4paper, oneside]{article}

\usepackage{latexsym}
\usepackage{amssymb}
\usepackage{mathrsfs}
\usepackage{graphicx}
\usepackage[latin1]{inputenc}
\usepackage{chicago}
\usepackage{multirow}
\usepackage{tabularx}
\usepackage{dcolumn}
\usepackage{amsthm}
\usepackage{amsmath}
\usepackage{amsfonts}
\usepackage{mathrsfs}
\usepackage[german]{varioref}
\usepackage{psfrag}
\usepackage{url}
\usepackage{rotating}
\usepackage{oldgerm}
\usepackage{algorithmic}
\usepackage{algorithm}
\usepackage{color}
\usepackage{footmisc}

\newcommand{\bfi}{\begin{fig}}
\newcommand{\efi}{\end{fig}}
\newcommand{\btab}{\begin{tabular}}
\newcommand{\etab}{\end{tabular}}
\newcommand{\barr}{\begin{array}}
\newcommand{\earr}{\end{array}}
\newcommand{\beqq}{\begin{equation}}
\newcommand{\eeqq}{\end{equation}}
\newcommand{\beao}{\begin{eqnarray*}}
\newcommand{\eeao}{\end{eqnarray*}\noindent}
\newcommand{\beam}{\begin{eqnarray}}
\newcommand{\eeam}{\end{eqnarray}\noindent}
\newcommand{\bdis}{\begin{displaymath}}
\newcommand{\edis}{\end{displaymath}\noindent}
\newcommand{\boldsymbolat}{\begin{pmatrix}}
\newcommand{\emat}{\end{pmatrix}}

\newtheorem{Theorem}{Theorem}[section]
\newtheorem{Example}[Theorem]{Example}
\newtheorem{Definition}[Theorem]{Definition}
\newtheorem{Lemma}[Theorem]{Lemma}
\newtheorem{Corollary}[Theorem]{Corollary}
\newtheorem{Proposition}[Theorem]{Proposition}

\newtheorem{Algorithm}[Theorem]{Algorithm}
\newtheorem{Remark}[Theorem]{Remark}

\numberwithin{equation}{section}

\newcolumntype{C}[1]{>{\centering\arraybackslash}p{#1}}

\long\def\symbolfootnote[#1]#2{\begingroup%
\def\thefootnote{\fnsymbol{footnote}}\footnote[#1]{#2}\endgroup}

\begin{document}

\title{Hierarchical Kendall copulas: Properties and inference}
\author{Eike Christian Brechmann\footnote{Center for Mathematical Sciences, Technische Universit\"at M\"unchen, Boltzmannstr. 3, D-85747 Garching, Germany. E-mail: eike.brechmann@tum.de. Phone: +49 89 289-17425. Fax: +49 89 289-17435.}}
\date{\today}
\maketitle

\begin{abstract}

While there is substantial need for dependence models in higher dimensions, most existing models quickly become rather restrictive and barely balance parsimony and flexibility.
Hierarchical constructions may improve on that by grouping variables in different levels.
In this paper, the new class of hierarchical Kendall copulas is proposed and discussed.
Hierarchical Kendall copulas are built up by flexible copulas specified for groups of variables, where aggregation is facilitated by the Kendall distribution function, the multivariate analog to the probability integral transform for univariate random variables. 
After deriving properties of the general model formulation, particular focus is given to inference techniques of hierarchical Kendall copulas with Archimedean components, for which closed-form analytical expressions can be derived.
A substantive application to German stock returns finally shows that hierarchical Kendall copulas perform very well for real data, out-of- as well as in-sample.

\end{abstract}

%---------------------------------------------------------%


%---------------------------------------------------------%

\section{Introduction}

Dependence modeling using copulas has made significant progress in the last years.
A $d$-dimensional copula is a multivariate distribution function on $[0,1]^d$ with uniform margins.
Their central role in dependence modeling is due to the famous theorem by \citeN{Sklar1959}, which states that every multivariate distribution function can be expressed in terms of a copula and the univariate marginal distribution functions.
More precisely, let $F_{\boldsymbol{X}}$ be the $d$-dimensional distribution function of the random vector $\boldsymbol{X}=(X_1,...,X_d)^\prime$ with univariate marginal distribution functions $F_{X_1},...,F_{X_d}$.
Then there exists a copula $C$ such that for all $\boldsymbol{x}=(x_1,...,x_d)^\prime\in[-\infty,\infty]^d$,
\begin{equation}
F_{\boldsymbol{X}}(\boldsymbol{x})=C(F_{X_1}(x_1),...,F_{X_d}(x_d)).
\label{eq:sklar}
\end{equation}
The copula $C$ is unique if $F_{X_1},...,F_{X_d}$ are continuous.

Many of the standard, and also of the newly proposed, copula models however turn out to be rather restrictive in higher dimensions, which makes it virtually impossible to use them for very large data sets as required, e.g., in financial or spatial applications.
While standard multivariate elliptical copulas such as the Gaussian and the Student's t require the specification of the full correlation matrix and can only account for symmetric dependence, multivariate Archimedean copulas are even more restrictive by assuming exchangeability and imposing that all multivariate margins are the same.
One common procedure to approach such problems therefore is grouping data, e.g., by industry sectors or nationality.
Such copula models include the grouped Student's t copula by \shortciteN{DaulDeGiorgiLindskogMcNeil2003}, elliptical copulas with clustered correlation matrix (see, e.g., \citeN{GregoryLaurent2004}) and hierarchical Archimedean copulas, which were initially proposed by \citeN{Joe1997}.
In particular, hierarchical structures such as the latter two are very appealing and received considerable attention lately (see, e.g., \citeN{Hofert2010}).

A major issue of any copula model is to find a good balance between parsimony and flexibility.
While elliptical copulas require an enormous number of parameters for specifying the correlation matrix (the number of parameters grows quadratically with the dimension), Archimedean and also hierarchical Archimedean copulas are much more parsimonious, since the number of parameters is at most linear in the dimension. 
However, such restrictions may be severe, since hierarchical Archimedean copulas are at the same time limited to the class of Archimedean copulas as building blocks.
Similarly, an elliptical copula with clustered correlation matrix has to satisfy positive definiteness constraints and is limited to an elliptical dependence structure, which in particular implies tail symmetry.

Another class of copulas that recently attracted increasing attention are pair copula constructions as proposed by \shortciteN{AasCzadoFrigessiBakken2009}.
Such pair copula constructions use bivariate copulas of arbitrary types as building blocks and are also available for applications in larger dimensions.
They however are non-hierarchical models and therefore may also severely suffer from extreme numbers of parameters.
While means to counteract these problems are now investigated \shortcite{BrechmannCzadoAas2012}, we focus here on hierarchical constructions, which are inherently more parsimonious and easier to interpret in terms of within- and between-group dependence.

The purpose of this paper is to introduce the new class of hierarchical Kendall copulas as a flexible, but yet parsimonious dependence model.
It is built up by copulas for groups (clusters) of variables in different hierarchical levels.
In particular---and in contrast to hierarchical Archimedean copulas---, the choice of copulas and their parameters is not restricted.
%it does not require any restrictions with respect to copula choice or the parameters.
With pair copula constructions the model shares the property that building blocks can be copulas of arbitrary types.
Hierarchical Kendall copulas therefore provide a new and attractive option to model dependence patterns between large numbers of variables.

The name ``hierarchical Kendall copula'' is chosen to stress the central role of the Kendall distribution function in the model formulation.
The Kendall distribution function is the multivariate analog to the probability integral transform for univariate random variables.
It is used to aggregate the (dependence) information of a group of variables.
It was first studied by \citeN{GenestRivest1993} in the bivariate case and in more detail by \shortciteN{BarbeGenestGhoudiRemillard1996}.
Other accounts on it can be found, amongst others, in \shortciteN{ImlahiEzzergChakak1999}, \citeN{GenestRivest2001} and \shortciteN{NelsenMolinaLallenaFlores2003} as well as in the copula goodness-of-fit literature (see, e.g., \citeN{WangWells2000} and \shortciteN{GenestQuessyRemillard2006}).

It has been shown by \shortciteN{GenestMolinaLallena1995} that the only copula that gives a valid multivariate distribution for non-overlapping multivariate marginals is the independence copula.
That is, if in Equation \eqref{eq:sklar} non-overlapping multivariate distribution functions instead of univariate ones are plugged into the copula $C$, this copula can only be the independence copula.
\citeN{MarcoRuizRivas1992} state conditions how a distribution function with specified multivariate marginals can be constructed; the easiest case being that margins are max-infinitely divisible, which includes distributions based on Archimedean copulas.
Hierarchical Kendall copulas circumvent such issues through aggregation facilitated by the Kendall distribution function.

The model, which we call hierarchical Kendall copula, has previously been mentioned by \citeN{AnjosKolev2005}, who however do not further develop the model in terms of statistical properties and inference.
The work presented here is completely independent of theirs and develops properties and inference techniques including sampling algorithms for hierarchical Kendall copulas, with particular focus on Archimedean cluster components.

The features of hierarchical dependence models and of hierarchical Kendall copulas in particular are attractive to different areas of applications.
In finance and insurance, risk capital needs to be aggregated over different levels of business lines and operating entities, which introduces a natural hierarchy with different dependencies across levels.
Also in other financial areas, there is a need for such models.
For instance, hierarchical Archimedean copulas have previously been used by \citeN{HofertScherer2011} for the pricing of CDOs.
For the purpose of market risk portfolio management, a substantial 30-dimensional application to German stock returns is presented in this paper, showing the need for careful dependence modeling and a good out-of- as well as in-sample performance of hierarchical Kendall copulas.

The model is however not limited to applications in finance and insurance, but may be used in any area that deals with some kind of clustered data such as geographic or temporal clusters. 
For instance, in hydrology Kendall distribution functions are used to characterize multivariate return periods (see \shortciteN{SalvadoriDeMicheleDurante2011}) and hierarchical Kendall copulas may be used to relate different return periods to each other.

The remainder of the paper is organized as follows.
The new model is introduced and discussed in Section \ref{sect:hiercop}.
Section \ref{sect:inference} then treats statistical inference techniques for hierarchical Kendall copulas, while the financial application is presented in Section \ref{sect:applic}.
Section \ref{sect:concl} concludes.

%---------------------------------------------------------%

\section{Hierarchical Kendall copulas}\label{sect:hiercop}

A central part of the definition of hierarchical Kendall copulas, which will be given below, is the notion of the Kendall distribution function, which is therefore treated first.
Subsequently, hierarchical Kendall copulas are defined and their properties are discussed, in particular in comparison to hierarchical Archimedean copulas.

\subsection{Kendall distribution functions}\label{sect:kendall}

Kendall distribution functions were first studied in two dimensions by \citeN{GenestRivest1993} and studied in more generality by \shortciteN{BarbeGenestGhoudiRemillard1996}.
For $\boldsymbol{U}:=(U_1,...,U_d)^\prime\sim C$, where $C$ is a $d$-dimensional copula, the Kendall distribution function $K^{(d)}$ is defined as
\begin{equation}
K^{(d)}(t) := P(C(\boldsymbol{U})\leq t),\quad t\in[0,1].
\label{eq:kendall}
\end{equation}
It holds that $t\leq K^{(d)}(t)\leq 1$ for $t\in[0,1]$ as well as $K^{(d)}(0-)=0$.

In this paper, it is assumed that copulas are absolutely continuous with continuous Kendall distribution functions.
Because the Kendall distribution function is the univariate distribution function of the random variable $Z:=C(\boldsymbol{U})$, we then have $K^{(d)}(Z)\sim U(0,1)$, which is the multivariate probability integral transform of $\boldsymbol{U}$.
Another interpretation is that $K^{(d)}$ describes the distribution of the level sets of a copula
\begin{equation}
L_C(z) = \{\boldsymbol{u}\in[0,1]^d: C(\boldsymbol{u})=z\},\quad z\in(0,1).
\label{eq:levelset}
\end{equation}

It has been shown by \citeN{GenestRivest1993} that bivariate Archimedean copulas are uniquely characterized by their Kendall distribution functions.
\shortciteN{GenestNeslehovaZiegel2011} recently extended this result to the trivariate case and strongly conjecture that this holds in general. 

The computation of the Kendall distribution function for a given copula is however complicated in general.
\shortciteN{ImlahiEzzergChakak1999} provide the recursive formula
\begin{equation}
K^{(d)}(t) = K^{(d-1)}(t) + \int_t^1 \int_{C_{u_1}^{-1}(t)}^1 ... \int_{C_{u_1,...,u_{d-2}}^{-1}(t)}^1 \int_0^{C_{u_1,...,u_{d-1}}^{-1}(t)} c(u_1,...,u_d)\, du_d...du_1,
\label{eq:kendallrecurs}
\end{equation}
where $K^{(d)}$ denotes the Kendall distribution function of the $d$-dimensional copula $C$ with density $c$ and $K^{(d-1)}$ that of the $(d-1)$-dimensional margin of the first $d-1$ variables.
Furthermore, the formula uses the notion of the copula quantile function, as studied by \shortciteN{ImlahiEzzergChakak1999} and \citeN{ChakakEzzerg2000}.
Define $C_{u_1,...,u_{d-1}}(\cdot) := C(u_1,...,u_{d-1},\cdot)$, then the copula quantile function is the inverse $C_{u_1,...,u_{d-1}}^{-1}$.
It holds that
\begin{equation*}
C(u_1,...,u_{d-1},C_{u_1,...,u_{d-1}}^{-1}(z)) = z
\end{equation*}
for $z\in(0,1)$.
For ease of notation, we define $C_{u_1,...,u_r}(\cdot) := C(u_1,...,u_r,\cdot,1,...,1)$ for $r=1,...,d-2$, and $C_{\emptyset}^{-1}(z) := z$ for $z\in(0,1)$.

Equation \eqref{eq:kendallrecurs} requires high-dimensional integration and availability of the copula quantile function in closed form. 
For general copulas, it is therefore not possible to easily determine the Kendall distribution function in closed form.
A convenient exception are however Archimedean copulas.
For a $d$-dimensional Archimedean copula (see \citeN{McNeilNeslehova2009}) with generator\footnote{The continuous and strictly decreasing function $\varphi: [0,1]\to[0,\infty)$ with $\varphi(1)=0$ generates a $d$-dimensional Archimedean copula if and only if its inverse $\varphi^{-1}$ is $d$-monotone on $[0,\infty)$, that is, $\varphi^{-1}$ is differentiable up to the order $d-2$ on $[0,\infty)$, it holds that $(-1)^k (\varphi^{-1})^{(k)}(x)\geq 0$ for $k=0,1,...,d-2$ and for any $x\in[0,\infty)$, and $(-1)^{d-2} (\varphi^{-1})^{(d-2)}$ is non-increasing and convex on $[0,\infty)$ (see \citeN{McNeilNeslehova2009}).} $\varphi$,
\begin{equation}
C(u_1,...,u_d) = \varphi^{-1}\left(\varphi(u_1)+...+\varphi(u_d)\right),
\label{eq:archcop}
\end{equation}
\shortciteN{BarbeGenestGhoudiRemillard1996} showed that the Kendall distribution function is given in terms of the generator $\varphi$ and higher order derivatives of its inverse $\varphi^{-1}$ as
\begin{equation}
K^{(d)}(t) = t + \sum_{i=1}^{d-1} \frac{(-1)^i}{i!}\varphi(t)^i(\varphi^{-1})^{(i)}(\varphi(t)),
\label{eq:archkendall}
\end{equation}
where $t\in(0,1]$.

\begin{Example}[Kendall distribution function]\label{ex:kendall}

Figure \ref{fig:Kindgumbel} shows Kendall distribution functions $K^{(d)}$ \eqref{eq:archkendall} of the independence copula ($\varphi(t)=-\log(t)$) and the Gumbel copula ($\varphi(t)=(-\log(t))^\theta$, $\theta\geq 1$) with parameter $\theta=2$ (corresponding to a Kendall's $\tau$ of 0.5) for different dimensions $d$.
Both Kendall distribution functions are, for fixed $t$, increasing with the dimension $d$.
The rate of increase in the case of the Gumbel copula with medium positive dependence is however much weaker.
The practical implications of this property on our model will be discussed in Section \ref{sect:modselect}.

\begin{figure}
\centering
\includegraphics[width=0.65\linewidth]{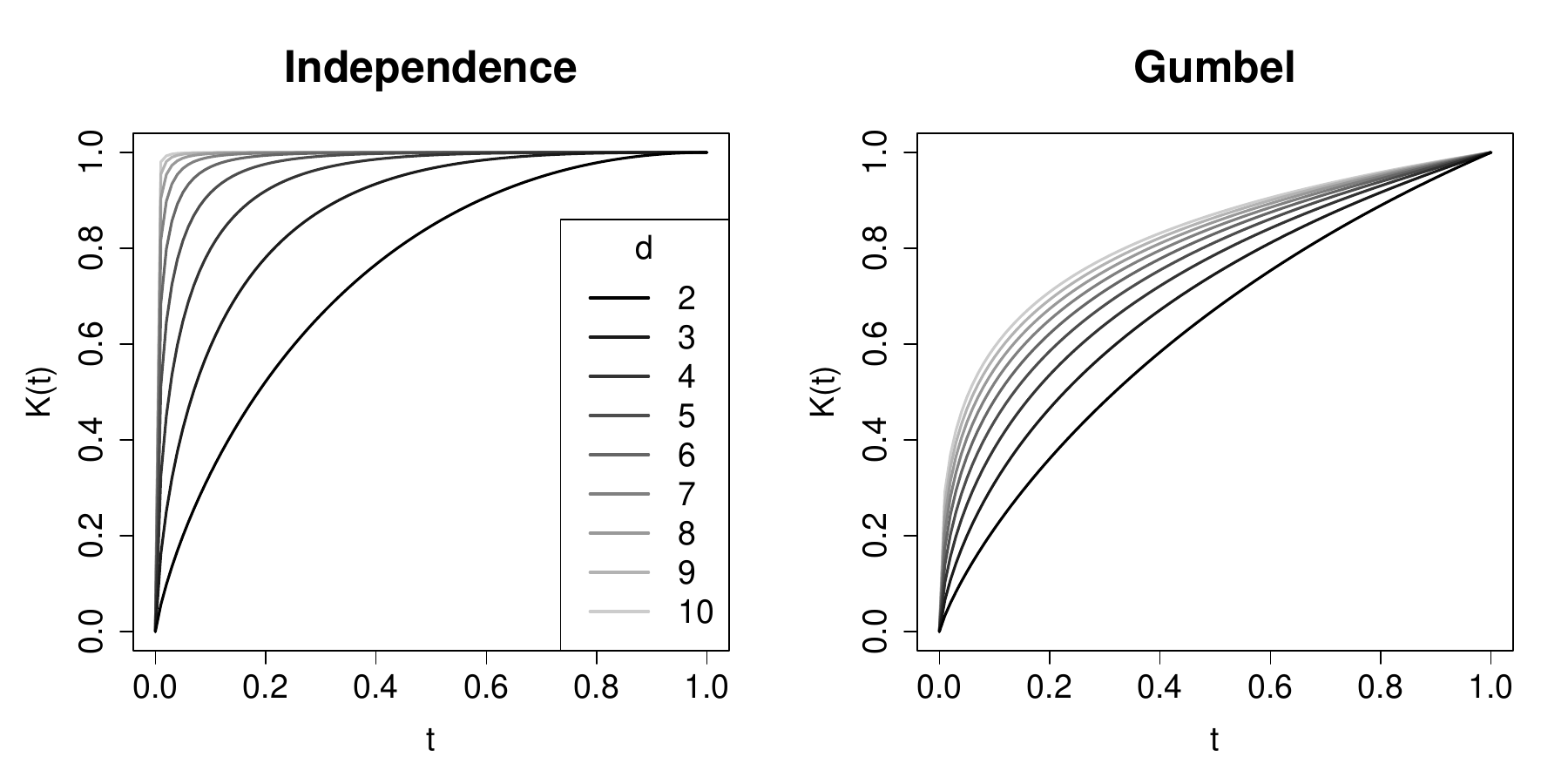}
\caption{Kendall distribution functions \eqref{eq:archkendall} of the independence copula (left panel) and the Gumbel copula with parameter $\theta=2$ (right panel) for different dimensions $d$.}
\label{fig:Kindgumbel}
\end{figure}

\end{Example}

In the following, the superscript of the Kendall distribution function, which indicates the dimension of the associated random vector, will be omitted for reasons of readability.

\subsection{Model formulation and properties}

The constructive definition of the new dependence model class of hierarchical Kendall copulas is now stated.
Although the model has previously been formulated by \citeN{AnjosKolev2005}, it has---to the best of our knowledge---not yet been treated in detail or used for statistical inference.
We choose the name ``hierarchical Kendall copula'' to stress the central role the Kendall distribution function plays in the model formulation.

\begin{Definition}[Hierarchical Kendall copula]\label{def:hiercop}

Let $U_1,...,U_n\sim U(0,1)$ and let $C_0,C_1,...,C_d$ be copulas of dimensions $d,n_1,...,n_d$, respectively, where $n_i\geq 1,\ i=1,...,d,$ and $n=\sum_{i=1}^d n_i$.
Further, let $K_1,...,K_d$ denote the Kendall distribution functions corresponding to $C_1,...,C_d$, respectively.
We define $m_i=\sum_{j=1}^{i} n_j$ for $i=1,...,d$, and $m_0=0$ as well as $\boldsymbol{U}_i:=(U_{m_{i-1}+1},...,U_{m_i})^\prime$ and $V_i:=K_i(C_i(\boldsymbol{U}_i))$ for $i=1,...,d$.
Under the assumptions that
\begin{itemize}

\item[$\mathcal{A}_1$:] $\boldsymbol{U}_1,...,\boldsymbol{U}_d$ are mutually independent conditionally on $(V_1,...,V_d)^\prime$, and

\item[$\mathcal{A}_2$:] the conditional distribution of $\boldsymbol{U}_i|(V_1,...,V_d)^\prime$ is the same as the conditional distribution of $\boldsymbol{U}_i|V_i$ for all $i=1,...,d$, that is, $F_{\boldsymbol{U}_i|V_1,...,V_d}=F_{\boldsymbol{U}_i|V_i}\ \forall i\in\{1,...,d\}$,

\end{itemize}
the random vector $(U_1,...,U_n)^\prime$ is said to be distributed according to the hierarchical Kendall copula $C_\mathcal{K}$ with nesting copula $C_0$ and cluster copulas $C_1,...,C_d$ if
\begin{enumerate}

\item $\boldsymbol{U}_i\sim C_i\ \forall i\in\{1,...,d\}$,

\item $(V_1,...,V_d)^\prime\sim C_0$.

\end{enumerate}

\end{Definition}

\begin{figure}[t]
\centering
\includegraphics[width=.9\linewidth]{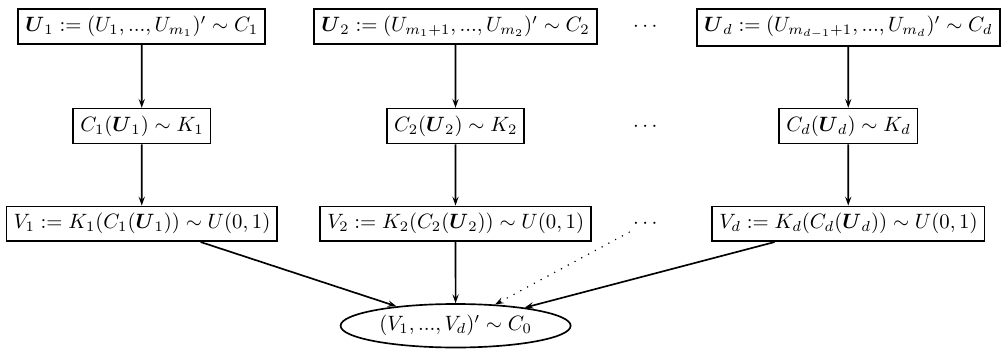}
\caption{Illustration of a hierarchical Kendall copula (see Definition \ref{def:hiercop}).}
\label{fig:hiercop}
\end{figure}

The distribution function $C_\mathcal{K}$ of $(U_1,...,U_n)^\prime$ will be characterized in terms of its density below.
First, we discuss the construction, which is illustrated in Figure \ref{fig:hiercop}, in more detail and provide examples.

The intuition behind the two assumptions $\mathcal{A}_1$ and $\mathcal{A}_2$ is that, given the information of the nesting variables $V_1,...,V_d$, the clusters $\boldsymbol{U}_1,...,\boldsymbol{U}_d$ are independent of each other and also of other nesting variables, since the dependence among the clusters is explained through the ``representatives'' $V_1,...,V_d$.
In other words, $V_1,...,V_d$ can be interpreted as unobserved factors, whose joint behavior determines the dependence of the different clusters.
In finance, such factors may be, e.g., industry sectors.

Note that $C_0$ is in general not the copula of $(U_1,...,U_n)^\prime$ but of $(V_1,...,V_d)^\prime$, which are uniform random variables due to $C_i(\boldsymbol{U}_i)\sim K_i$ for all $i=1,...,d$.
The nesting copula $C_0$ and the cluster copulas $C_1,...,C_d$ can be chosen independently.
They can be arbitrary copulas such as common Archimedean or elliptical copulas or from any other class of copulas.
A special case of hierarchical Kendall copulas is the upper Fr\'echet-Hoeffding bound.

\begin{Example}[Upper Fr\'echet-Hoeffding bound]\label{rem:comon}

Let $C_\mathcal{K}$ be a hierarchical Kendall copula, where the clusters are perfectly positively dependent and the nesting copula also is the upper Fr\'echet-Hoeffding bound $C_0(v_1,...,v_d)=M^{(d)}(v_1,...,v_d):=\min\{v_1,...,v_d\}$.
Since $C_i=M^{(n_i)}$ and $K_i(t)=t$ for all $i=1,...,d$ and $t\in[0,1]$, it holds that $V_i=\min\{U_{m_{i-1}+1},...,U_{m_i}\}$ and therefore $C_\mathcal{K}=M^{(n)}$.
In other words, the upper Fr\'echet-Hoeffding bound belongs to the class of hierarchical Kendall copulas.

\end{Example}

The Kendall distribution functions $K_1,...,K_d$ are used to summarize the information contained in the clusters (transformation of $n_i$-variate to univariate random vector).
While one may also think of other transformations, we believe that Kendall distribution functions are particularly useful for this purpose, since they have a range of  properties that are reasonable to demand of an aggregating function.
The properties mainly follow from the definitions of a copula and the Kendall distribution function \eqref{eq:kendall}.

\begin{Proposition}[Properties of aggregation with Kendall distribution functions]\label{prop:kendallagg}

Let $K$ be the Kendall distribution function of a $p$-dimensional copula $C$.
Then the aggregation function $K\circ C: [0,1]^p \to [0,1], \boldsymbol{u}\mapsto K(C(\boldsymbol{u}))$ has the following properties:

\begin{enumerate}

\item The function $K\circ C$ is monotone increasing, that is,
\begin{equation*}
K\circ C(u_1,...,u_p) \leq K\circ C(v_1,...,v_p)
\end{equation*}
for every $u_1,...,u_p,v_1,...v_p\in[0,1]$ such that $u_i\leq v_i,\ i=1,...,p$.

\item If $K(0)=0$, then $K\circ C$ is grounded, that is, $K\circ C(u_1,...,u_p) = 0$ if any $u_i=0,\ i\in\{1,...,p\}$, in particular $K\circ C(0,...,0) = 0$.

\item It holds that $K\circ C(1,...,1) = 1$.

\item In the degenerate case that $p=1$, the function $K\circ C$ reduces to the identity function, that is, $K\circ C(u_1)=u_1$ for $u_1\in[0,1]$.

\item If $C$ is exchangeable, then $K\circ C$ is commutative, that is, 
\begin{equation*}
K\circ C(u_1,...,u_p) = K\circ C(u_{\pi(1)},...,u_{\pi(p)})
\end{equation*}
for every $u_1,...,u_p\in[0,1]$ and any permutation $\pi: \{1,...,p\}\to\{1,...,p\}$.

\end{enumerate}

\end{Proposition}

The condition $K(0)=0$ in the second property is fulfilled in particular by all absolutely continuous copulas.
Examples of exchangeable copulas are Archimedean \eqref{eq:archcop} and elliptical copulas with equi-correlation matrix.

The nesting copula $C_0$ then essentially models the comovement of the cluster copula level sets \eqref{eq:levelset}. 
This can be seen as a proxy for the strength of dependence in the clusters, since the dimensionality of the single clusters is ``normalized'' through the Kendall distribution functions.
Other variables such as components of an elliptical distribution do not contain such specific information that summarizes the information in a single variable.
%and a focus on tails, as it is often desired, is empirically hardly feasible, since tail behavior is very hard to quantify appropriately.
Thus, the transformation using the Kendall distribution function reasonably summarizes the (dependence) information of a multivariate random vector in the spirit of the univariate probability integral transform and with the appealing characteristics stated in Proposition \ref{prop:kendallagg}.
For the particular purpose of risk aggregation, an alternative hierarchical dependence model for sums of random variables was recently proposed by \shortciteN{ArbenzHummelMainik2011}.

In the case of Archimedean clusters, we can give a particularly convenient representation of the multivariate distribution of $U_1,...,U_n$.

\begin{Remark}[Hierarchical Kendall copula with Archimedean clusters]\label{rem:archclust}

Let $\boldsymbol{U}:=(U_1,...,U_n)^\prime$ be distributed according to a hierarchical Kendall copula $C_\mathcal{K}$, where $C_1,...,C_d$ are Archimedean with generators $\varphi_1,...,\varphi_d$, respectively.
According to \citeN{McNeilNeslehova2009}, it holds for all $i=1,...,d$,
\begin{equation}
(\varphi_i(U_{m_{i-1}+1}),...,\varphi_i(U_{m_i}))^\prime \stackrel{d}{=} R_i\boldsymbol{S}^{(i)},
\label{eq:arch}
\end{equation}
where $\boldsymbol{S}^{(i)}=(S_1^{(i)},...,S_{n_i}^{(i)})^\prime$ is uniformly distributed on the unit simplex $\{\boldsymbol{x}\geq 0: \sum_{j=1}^{n_i} x_j = 1\}\subset[0,1]^{n_i}$, and the radial part $R_i=\sum_{j=1}^{n_i} \varphi_i(U_{m_{i-1}+j})$ is independent of $\boldsymbol{S}^{(i)}$ and
has distribution $F_{R_i}$, which can be determined through the inverse Williamson transform of $\varphi_i^{-1}$.

As a result we can represent the random vector $\boldsymbol{U}$ as
\begin{equation}
(U_1,...,U_n)^\prime \stackrel{d}{=} (\varphi_1^{-1}(R_1S_1^{(1)}),...,\varphi_1^{-1}(R_1S_{n_1}^{(1)}),\varphi_2^{-1}(R_2S_1^{(2)}),...,\varphi_2^{-1}(R_2S_{n_2}^{(2)}),...,\varphi_d^{-1}(R_dS_{n_d}^{(d)}))^\prime,
\label{eq:archclust}
\end{equation}
where $R_i=\varphi_i(K_i^{-1}(V_i))$ for $i=1,...,d$, since by the definition of $C_i$, $V_i$ and $R_i$,
\begin{equation*}
V_i=K_i(C_i(\boldsymbol{U}_i))=K_i\Big(\varphi_i^{-1}\Big(\sum_{j=1}^{n_i} \varphi_i(U_{m_{i-1}+j})\Big)\Big)=K_i(\varphi_i^{-1}(R_i)).
\end{equation*}
In other words, if all clusters are Archimedean, dependence among clusters is introduced solely through the dependence between the radial variables of the different clusters.
In particular, if the nesting copula $C_0$ is also Archimedean with generator $\varphi_0$ and corresponding radial variable $R_0$, we have for $i=1,...,d$ that $V_i=\varphi_0^{-1}(R_0S_{i}^{(0)})$, where $S_1^{(0)},...,S_{d}^{(0)}$ are uniformly distributed on the $d$-dimensional unit simplex.
That is, the radial variables of the clusters, $R_i$, can be expressed through $R_0$ and uniform random variables on the simplex.

Equation \eqref{eq:archclust} also motivates to speak of a ``grouped Archimedean copula'' similar to the grouped Student's t copula by \shortciteN{DaulDeGiorgiLindskogMcNeil2003}.

\end{Remark}

We now provide an illustrative example of a hierarchical Kendall copula with Archimedean clusters.

\begin{Example}[Hierarchical Kendall copula with Archimedean clusters]\label{ex:archclust}

Let $C_\mathcal{K}$ be a four-dimensional hierarchical Kendall copula with $n_1=n_2=2$.
The bivariate cluster copulas are chosen as Clayton (Archimedean with $\varphi(t)=t^{-\theta}-1$, $\theta>0$) with parameter $\theta=1.33$ and Gumbel with parameter $\theta=1.67$ (both corresponding to a Kendall's $\tau$ of $0.4$). 
The nesting copula is set as a Frank (also Archimedean with $\varphi(t)=-\log((1-e^{-\theta t})/(1-e^{-\theta}))$, $\theta\in\mathbb{R}\setminus \{0\}$) with parameter $\theta=11.41$ (Kendall's $\tau$ of $0.7$).
A sample of size 1000 from this hierarchical Kendall copula is shown in Figure \ref{fig:exarch}.
It shows the typical features of lower tail dependence for the pair $(U_1,U_2)^\prime$ (Clayton copula) and of upper tail dependence for the pair $(U_3,U_4)^\prime$ (Gumbel copula).
The between-cluster dependence looks rather tail-symmetric as implied by the Frank copula.

\begin{figure}[t]
\centering
\includegraphics[width=.7\linewidth]{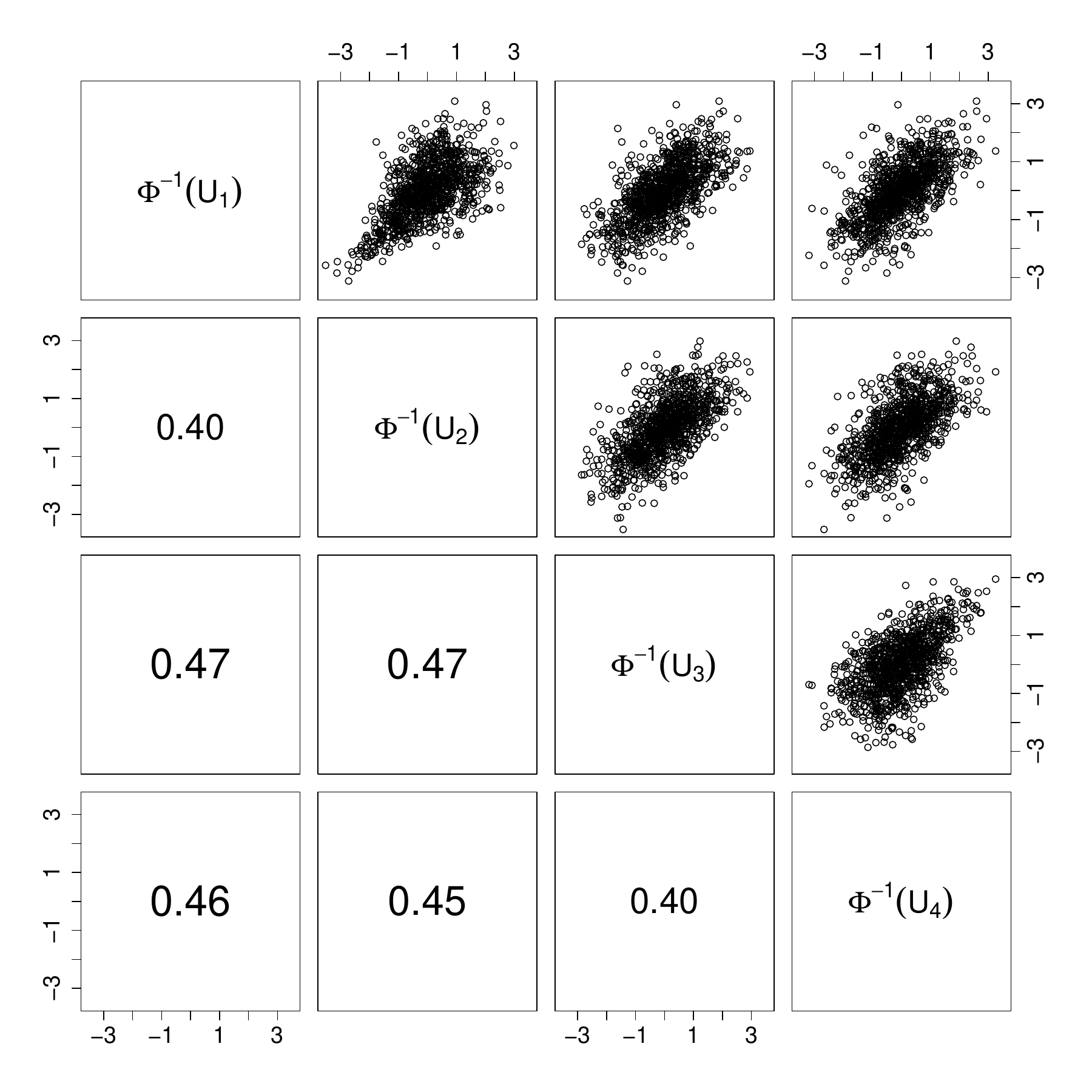}
\caption{A sample of size 1000 from a four-dimensional hierarchical Kendall copula with Clayton and Gumbel clusters and Frank nesting copula. Margins are transformed to standard normal. The lower triangle shows pairwise empirical Kendall's $\tau$ values.}
\label{fig:exarch}
\end{figure}

\end{Example}

The illustration in Example \ref{ex:archclust} also provides an example where the between-cluster dependence is stronger than the within-cluster dependence.
This case cannot be modeled using hierarchical Archimedean copulas as will be discussed in Section \ref{sect:hierarch}.

The two-level construction given in Definition \ref{def:hiercop} can also be extended to an arbitrary number of levels.

\begin{Remark}[Hierarchical Kendall copula with $k$ levels]\label{def:klevelhiercop}

Let $U_1,...,U_n\sim U(0,1)$ and let $d_j,\ j=1,...,k-1,$ denote the number of clusters per level $j$, such that $d_1\geq d_2\geq...\geq d_{k-1}$.
Further, let the nesting copula $C_0$ be $d_{k-1}$-dimensional and let the nested cluster copulas $C_i^{(j)}$, $j=1,...,k-1,\ i=1,...,d_j$, be of dimension $n_i^{(j)}\geq 1$, where $n=\sum_{i=1}^{d_1} n_i^{(1)}$ and $d_{j-1}=\sum_{i=1}^{d_j} n_i^{(j)}$ for $j=2,...,k-1$.
The index $i$ runs over the $d_j$ clusters of each level $j$.
The Kendall distribution function corresponding to $C_i^{(j)},\ j=1,...,k-1,\ i=1,...,d_j,$ is denoted by $K_i^{(j)}$ and we define $m_i^{(j)}=\sum_{\ell=1}^{i} n_\ell^{(j)}$ for $i=1,...,d_j$, and $m_0^{(j)}=0$.
Under independence assumptions as in Definition \ref{def:hiercop}, we then say that the random vector $(U_1,...,U_n)^\prime$ is distributed according to the $k$-level hierarchical Kendall copula $C_\mathcal{K}$ with nesting copula $C_0$ and cluster copulas $C_i^{(j)},\ j=1,...,k-1,\ i=1,...,d_j,$ if
\begin{enumerate}

\item $\boldsymbol{U}_i:=(U_{m_{i-1}^{(1)}+1},...,U_{m_i^{(1)}})^\prime\sim C_i^{(1)}\ \forall i\in\{1,...,d_1\}$,

\item $V_i^{(1)}:=K_i^{(1)}(C_i^{(1)}(\boldsymbol{U}_i))\  \forall i\in\{1,...,d_1\}$,

\item for $j=2,...,k-1$:

\begin{enumerate}

\item $\boldsymbol{V}_i^{(j-1)}:=(V_{m_{i-1}^{(j)}+1}^{(j-1)},...,V_{m_i^{(j)}}^{(j-1)})^\prime\sim C_i^{(j)}\ \forall i\in\{1,...,d_j\}$,

\item $V_i^{(j)}:=K_i^{(j)}(C_i^{(j)}(\boldsymbol{V}_i^{(j-1)}))\  \forall i\in\{1,...,d_j\}$,

\end{enumerate}

\item $(V_1^{(k-1)},...,V_{d_{k-1}}^{(k-1)})^\prime\sim C_0$.

\end{enumerate}
In particular, the clusters $\boldsymbol{U}_1,...,\boldsymbol{U}_{d_1}$ at the lowest level ($j=1$) are assumed to be mutually independent given the ``representatives'' $V_i^{(j)},\ j=1,...,k-1,\ i=1,...,d_j$.
In the case of $k=3$, we define
\begin{equation*}
\boldsymbol{U}_i^{(1)}:=(\boldsymbol{U}_{m_{i-1}^{(2)}+1}^\prime,...,\boldsymbol{U}_{m_i^{(2)}}^\prime)^\prime,\quad i=1,...,d_2.
\end{equation*}
Then the assumptions are
\begin{itemize}

\item[$\mathcal{A}_1^{(1)}$:] $\boldsymbol{U}_1^{(1)},.., \boldsymbol{U}_{d_2}^{(1)}$ are mutually independent conditionally on $(V_1^{(2)},...,V_{d_2}^{(2)})^\prime$;

\item[$\mathcal{A}_2^{(1)}$:] the conditional distribution of  $\boldsymbol{U}_i^{(1)}|(V_1^{(2)},...,V_{d_2}^{(2)})^\prime$ is the same as the conditional distribution of $\boldsymbol{U}_i^{(1)}|V_i^{(2)}$ for all $i=1,...,d_2$;

\item[$\mathcal{A}_1^{(2)}$:] $\boldsymbol{U}_{m_{i-1}^{(2)}+1},..., \boldsymbol{U}_{m_{1}^{(2)}}$ are mutually independent conditionally on $(V_i^{(2)},\boldsymbol{V}_i^{(1)\prime})^\prime$ for all $i=1,...,d_2$;

\item[$\mathcal{A}_2^{(2)}$:] the conditional distribution of $\boldsymbol{U}_{m_{i-1}^{(2)}+j}|(V_i^{(2)},\boldsymbol{V}_i^{(1)\prime})^\prime$ is the same as the conditional distribution of $\boldsymbol{U}_{m_{i-1}^{(2)}+j}|(V_i^{(2)},V_{m_{i-1}^{(2)}+j}^{(1)})^\prime$ for all $i=1,...,d_2,\ j=1,...,n_i^{(2)}$.

\end{itemize}
Their interpretation is essentially the same as for the assumptions $\mathcal{A}_1$ and $\mathcal{A}_2$ stated in Definition \ref{def:hiercop}.

\end{Remark}

\begin{figure}[t]
\centering
\includegraphics[width=\linewidth]{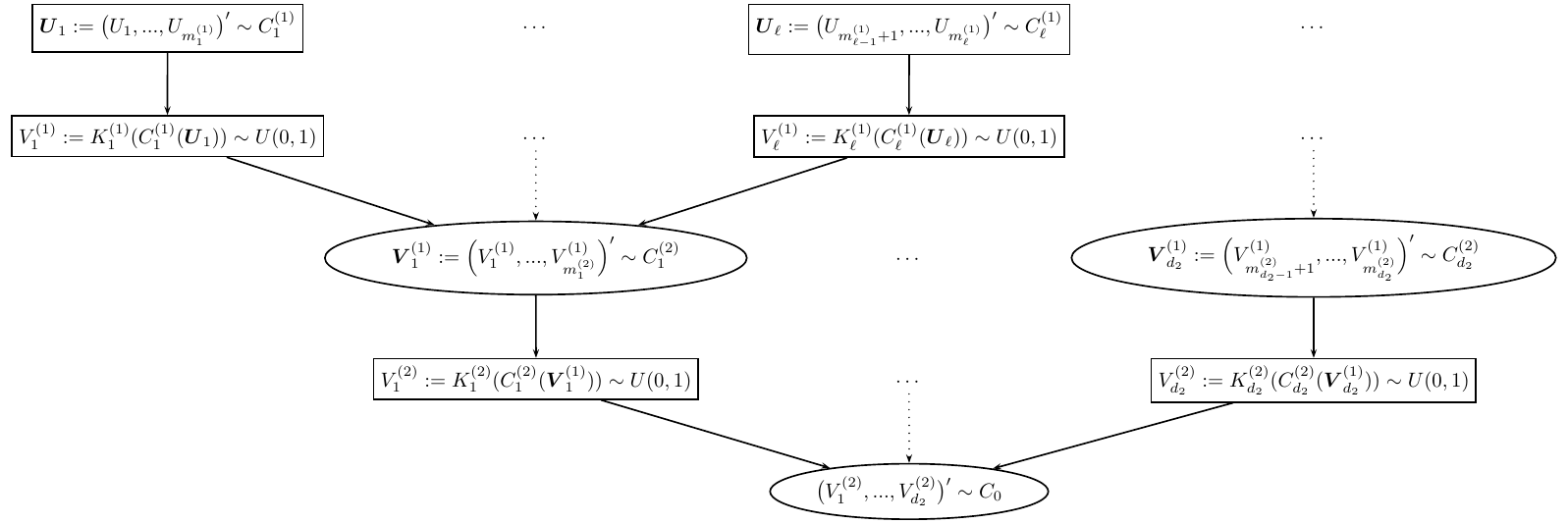}
\caption{Illustration of a three-level hierarchical Kendall copula (see Remark \ref{def:klevelhiercop}). Here $m_1^{(2)}=\ell$.}
\label{fig:hiercop2}
\end{figure}

An example of a three-level hierarchical Kendall copula is shown in Figure \ref{fig:hiercop2}.
For simplicity and illustrative reasons, we restrict our exposition here to the case of $k=2$ hierarchical levels.
It will be sketched how to generalize all derivations and methods, which are described in the following, to the general $k$ level case.

The two independence assumptions $\mathcal{A}_1$ and $\mathcal{A}_2$ of Definition \ref{def:hiercop} provide a natural structure for hierarchical dependence models and allow to derive the joint density function of a hierarchical Kendall copula as stated in the following theorem.
Densities of the copulas $C_0,...,C_d$ are denoted by $c_0,...,c_d$, respectively.

\begin{Theorem}[Joint density of a hierarchical Kendall copula]\label{theo:joint}

Let $\boldsymbol{U}=(U_1,...,U_n)^\prime$ be distributed according to a hierarchical Kendall copula $C_\mathcal{K}$ with nesting copula $C_0$ and cluster copulas $C_1,...,C_d$. 
The joint density function $c_{\mathcal{K}}$ of $C_\mathcal{K}$ is then given as follows:
\begin{equation} 
c_{\mathcal{K}}(\boldsymbol{u}) = c_0(K_1(C_1(\boldsymbol{u}_1)),...,K_d(C_d(\boldsymbol{u}_d))) \prod_{i=1}^d c_i(\boldsymbol{u}_i),
\label{eq:jointdens}
\end{equation}
where $\boldsymbol{u}=(u_1,...,u_n)^\prime$ and $\boldsymbol{u}_i=(u_{m_{i-1}+1},...,u_{m_i})^\prime,\ i=1,...,d$.

\end{Theorem}

\noindent
\textbf{Proof:} It is
\begin{equation}
C_{\mathcal{K}}(\boldsymbol{u}) = P(\boldsymbol{U}\leq\boldsymbol{u}) = \int_{[0,1]^d} P(\boldsymbol{U}\leq\boldsymbol{u}|V_1=v_1,...,V_d=v_d)\, c_0(v_1,...,v_d)\, dv_1...dv_d.
\label{eq:jointdens1}
\end{equation}
By applying assumption $\mathcal{A}_1$ first and then assumption $\mathcal{A}_2$, we obtain
\begin{equation*}
\begin{split}
P(\boldsymbol{U}\leq\boldsymbol{u}|V_1=v_1,...,V_d=v_d) & = \prod_{i=1}^d P(\boldsymbol{U}_i\leq\boldsymbol{u}_i|V_1=v_1,...,V_d=v_d)\\
& = \prod_{i=1}^d P(\boldsymbol{U}_i\leq\boldsymbol{u}_i|V_i=v_i).
\end{split}
\end{equation*}
Using this result, Equation \eqref{eq:jointdens1} simplifies to
\begin{equation}
C_{\mathcal{K}}(\boldsymbol{u}) = \int_{[0,1]^d} \left[ \prod_{i=1}^d F_{\boldsymbol{U}_i|V_i}(\boldsymbol{u}_i|v_i)\right]c_0(v_1,...,v_d)\, dv_1...dv_d.
\label{eq:jointdens1a}
\end{equation}
Further, we denote by $\boldsymbol{U}_{i,-m_i},\ i\in\{1,...,d\},$ the $(n_i-1)$-dimensional sub-vector of $\boldsymbol{U}_i=(U_{m_{i-1}+1},...,U_{m_i})^\prime$ with element $U_{m_i}$ removed.
Since $V_i=K_i(C_i(\boldsymbol{U}_i))\sim U(0,1)$ and according to a change of variables, it then holds that
\begin{equation}
\begin{split}
f_{\boldsymbol{U}_{i,-m_i}|V_i}(\boldsymbol{u}_{i,-m_i}|v_i) &= f_{\boldsymbol{U}_{i,-m_i},V_i}(\boldsymbol{u}_{i,-m_i},v_i)\\
&= c_i(\boldsymbol{u}_{i,-m_i},C_{i;\boldsymbol{u}_{i,-m_i}}^{-1}(K_i^{-1}(v_i))) \frac{\partial}{\partial v_i} C_{i;\boldsymbol{u}_{i,-m_i}}^{-1}(K_i^{-1}(v_i)),
\end{split}
\label{eq:jointdens1b}
\end{equation}
if $v_i\leq K_i(C_i(\boldsymbol{u}_{i,-m_i},1))$.
This yields
\begin{equation*}
F_{\boldsymbol{U}_i|V_i}(\boldsymbol{u}_i|v_i) = \int_0^{u_{m_i-1}}...\int_0^{u_{m_{i-1}+1}} f_{\boldsymbol{U}_{i,-m_i}|V_i}(\boldsymbol{w}_{i,-m_i}|v_i)\, 1_{\{C_{i;\boldsymbol{w}_{i,-m_i}}^{-1}(K_i^{-1}(v_i))\leq u_{m_i}\}}\, d\boldsymbol{w}_{i,-m_i}.
\end{equation*}
Plugging this expression into Equation \eqref{eq:jointdens1a} and substituting $v_i$ by $w_{m_i}=C_{i;\boldsymbol{w}_{i,-m_i}}^{-1}(K_i^{-1}(v_i))$ for $i=1,...,d$ then leads to
\begin{equation*}
C_{\mathcal{K}}(\boldsymbol{u}) = \int_0^{u_n}...\int_0^{u_1} \left[ \prod_{i=1}^d c_i(\boldsymbol{w}_i)\right] c_0(K_1(C_1(\boldsymbol{w}_1)),...,K_d(C_d(\boldsymbol{w}_d)))\, d\boldsymbol{w},
\end{equation*}
where we used that $v_i=K_i(C_i(\boldsymbol{w}_i)),\ i=1,...,d$.
Taking derivatives with respect to $u_1,...,u_n$ therefore gives the desired result.\hfill$\square$\\

%By the definition of $V_i=K_i(C_i(\boldsymbol{U}_i))$, the random vector $(\boldsymbol{U}_i,V_i)$ only has positive mass at $(\boldsymbol{u}_i,v_i)$ if $\boldsymbol{u}_i$ lies in the level set $L_{K_i\circ C_i}(v_i)=\{\boldsymbol{u}_i\in[0,1]^{n_i}: K_i(C_i(\boldsymbol{u}_i))=v_i\}$.
%Then it holds that $v_i=K_i(C_i(\boldsymbol{u}_i))$ and the mass at $(\boldsymbol{u}_i,v_i)$ is $c_i(\boldsymbol{u}_i)$.
%Taking derivatives of Equation \eqref{eq:jointdens1a} with respect to $u_1,...,u_n$ therefore gives the desired result.\hfill$\square$\\

\begin{Remark}[Joint density of a $k$-level hierarchical Kendall copula]\label{rem:jointdensklevel}

The arguments of Theorem \ref{theo:joint} can be iterated to derive the joint density of a $k$-level hierarchical Kendall copula.
By first conditioning on the aggregated variables of level $k-1$, $V_1^{(k-1)},...,V_{d_{k-1}}^{(k-1)}$, then on those of level $k-2$ up to level 1, an expression similar to Equation \eqref{eq:jointdens} is obtained.

For instance, the density of the three-level hierarchical Kendall copula can be derived along the lines of the proof of Theorem \ref{theo:joint} as
\begin{equation}
\begin{split}
c_\mathcal{K}(\boldsymbol{u}) &= c_0(v_1^{(2)},...,v_{n_2}^{(2)}) \prod_{i=1}^{d_2} \left( c_i^{(2)}\left(\boldsymbol{v}_i^{(1)}\right) \prod_{j=1}^{n_i^{(2)}} c_{m_{i-1}^{(2)}+j}^{(1)}\left(\boldsymbol{u}_{m_{i-1}^{(2)}+j}\right) \right)\\
&= c_0(v_1^{(2)},...,v_{n_2}^{(2)}) \prod_{i=1}^{d_2} c_i^{(2)}\left(\boldsymbol{v}_i^{(1)}\right) \prod_{i=1}^{d_1} c_i^{(1)}\left(\boldsymbol{u}_i\right),
\end{split}
\label{eq:jointdens3}
\end{equation}
where $\boldsymbol{v}_i^{(1)}=(v_{m_{i-1}^{(2)}+1}^{(1)},...,v_{m_i^{(2)}}^{(1)})^\prime,\ i=1,...,d_2,$ with components $v_i^{(1)}=K_i^{(1)}(C_i^{(1)}(\boldsymbol{u}_i)),\ i=1,...,d_1$.
Further, $v_i^{(2)}=K_i^{(2)}(C_i^{(2)}(\boldsymbol{v}_i^{(1)})),\ i=1,...,d_2$.

This means that the density of a three-level hierarchical Kendall copula also conveniently decomposes into the product of copula densities, where the arguments are obtained through the repeated application of Kendall distribution functions.

\end{Remark}

Another important special case of hierarchical Kendall copulas can easily be stated using Theorem \ref{theo:joint}.

\begin{Example}[Independence copula]\label{rem:indep}

Let $C_\mathcal{K}$ be a hierarchical Kendall copula, where both cluster and nesting copulas are independence copulas.
Since the independence copula has density equal to 1, it follows that $c_{\mathcal{K}}(\boldsymbol{u})=1$.
This means that the independence copula also belongs to the class of hierarchical Kendall copulas.

\end{Example}

Theorem \ref{theo:joint} also allows to formulate the following corollary which summarizes the marginal properties of hierarchical Kendall copulas.

\begin{Corollary}[Margins of a hierarchical Kendall copula]\label{cor:marg}

The same notation as in Theorem \ref{theo:joint} is used.

\begin{enumerate}

\item Bivariate margins: Let $k,\ell\in\{1,...,n\},\ k\neq\ell$. W.l.o.g. $k<\ell$.

\begin{enumerate}

\item If $U_k$ and $U_\ell$ are in the same cluster $i$, their marginal distribution function $C_{\mathcal{K},k\ell}$ is
\begin{equation*}
C_{\mathcal{K},k\ell} (u_k,u_\ell) := C_i(1,...,1,u_k,1,...,1,u_\ell,1,...,1).
\end{equation*}

\item If $U_k$ and $U_\ell$ are in different clusters $i$ and $j$, respectively, their marginal distribution function $C_{\mathcal{K},k\ell}$ is
\begin{equation}
\begin{split}
C_{\mathcal{K},k\ell}(u_k,u_\ell) := \int_0^{u_k}\int_0^{u_\ell} \int_{[0,1]^{n_i+n_j-2}} &
c_{0,ij}(K_i(C_i(\boldsymbol{w}_i)),K_j(C_j(\boldsymbol{w}_j))) \\
& \times c_i(\boldsymbol{w}_i)\, c_j(\boldsymbol{w}_j)\, d\boldsymbol{w}_{i,-k}\, d\boldsymbol{w}_{j,-\ell}\, dw_\ell\, dw_k,
\end{split}
\label{eq:bivmarg}
\end{equation}
where $c_{0,ij}$ is the density of the bivariate $(i,j)$-margin of $C_0$. 
%and $\boldsymbol{w}_{i,-k}$ denotes the $(n_i-1)$-dimensional sub-vector of $\boldsymbol{w}_i=(w_{m_{i-1}+1},...,w_{m_i})^\prime$ with element $w_k$ removed.
%Similarly for $\boldsymbol{w}_{j,-\ell}$.

\end{enumerate}

\item Multivariate margins: The marginal distribution of the cluster $\boldsymbol{U}_i$ is $C_i$.

\end{enumerate}

\end{Corollary}

More general multivariate margins involving variables from different clusters can be derived as in Equation \eqref{eq:bivmarg}.

\begin{Remark}[Mixture representation]

As a consequence of Corollary \ref{cor:marg} $(i)(b)$, bivariate marginal distributions where the variables are in different clusters can be regarded as a kind of continuous mixture of the nesting copula $C_0$.
The density of $C_{\mathcal{K},k\ell}$ as defined above is given by
\begin{equation*}
c_{\mathcal{K},k\ell} (u_k,u_\ell) = \int_{[0,1]^{n_i+n_j-2}}
c_{0,ij}(K_i(C_i(\boldsymbol{u}_i)),K_j(C_j(\boldsymbol{u}_j)))\, c_i(\boldsymbol{u}_i)\, c_j(\boldsymbol{u}_j)\, d\boldsymbol{u}_{i,-k}\, d\boldsymbol{u}_{j,-\ell},
\end{equation*}
where the mixing density weights are given by the product $c_i(\boldsymbol{u}_i)c_j(\boldsymbol{u}_j)$.

This representation complements the results of Remark \ref{rem:archclust}.
Equation \eqref{eq:archclust} shows that hierarchical Kendall copulas with Archimedean cluster copulas can be represented as transformed mixtures of uniform distributions on unit simplices.

\end{Remark}

As noted above, the multivariate distribution of $U_1,...,U_n$ defined through a hierarchical Kendall copula is in general not the copula $C_0$ but given through its density (see Equation \eqref{eq:jointdens1}).
We showed that the important special cases of independence as well as of comonotonicity are hierarchical Kendall copulas (see Examples \ref{rem:comon} and \ref{rem:indep}), while in general dependence between clusters ranges between these cases and can also be negative.
It is yet an open question which other common multivariate distributions can be represented as hierarchical Kendall copulas with non-trivial cluster sizes (at least one $n_i>1$, that is, $d<n$).
In particular, hierarchical Archimedean copulas are different from hierarchical Kendall copulas---with positive and negative implications as discussed next.

\subsection{Comparison with hierarchical Archimedean copulas}\label{sect:hierarch}

The popular class of hierarchical Archimedean copulas also allows for a nested modeling of clusters of variables (see, e.g., \citeN{Joe1997}, \citeN{Hofert2010}, \citeN{SavuTrede2010} and \shortciteN{OkhrinOkhrinSchmid2012}).
In contrast to hierarchical Kendall copulas, hierarchical Archimedean copulas are however limited to Archimedean copulas as building blocks, while hierarchical Kendall copulas can be built up by any possible copula.
Furthermore, hierarchical Archimedean copulas require stronger within-cluster dependence, which results in parameter restrictions if generators are the same.
This is not the case for hierarchical Kendall copulas (see Example \ref{ex:archclust}).

Archimedean copulas have many useful properties.
For instance, their relationship to Laplace transforms can be used to conveniently sample from hierarchical Archimedean copulas as described, e.g., in \citeN[2011]{Hofert2010}\nocite{Hofert2011}.
Hierarchical Kendall copulas also benefit from many of these properties.
It will be shown in Section \ref{sect:sim} that closed-form sampling of hierarchical Kendall copulas is feasible, when cluster copulas are Archimedean.
Hierarchical Kendall copulas with Archimedean cluster copulas are further particularly easy to estimate, since Kendall distribution functions are known in closed form for Archimedean copulas.
For that reason they also provide a closed-form joint density function, which is numerically tractable even in higher dimensions (see Equation \eqref{eq:jointdens3} for the case of three hierarchical levels).
Although the general density expression of hierarchical Archimedean copulas is in general hardly accessible (see \citeN{SavuTrede2010}), \citeN{HofertPham2012} recently derived a tractable formula for the case of a moderate number of nesting levels.

In contrast to hierarchical Archimedean copulas, multivariate margins of hierarchical Kendall copulas are however not directly available (see Corollary \ref{cor:marg}).
For instance, if $(U_1,U_2)^\prime\sim C_1$, $(U_3,U_4)^\prime\sim C_2$ and the nesting copula in both hierarchical copula models is $C_0$, then in the hierarchical Archimedean copula the margin $(U_1,U_3)^\prime$ is distributed according to $C_0$, while in the case of the hierarchical Kendall copula this marginal distribution has to be obtained using integration as in Equation \eqref{eq:bivmarg}.

\begin{figure}
\centering
\includegraphics[width=0.325\linewidth]{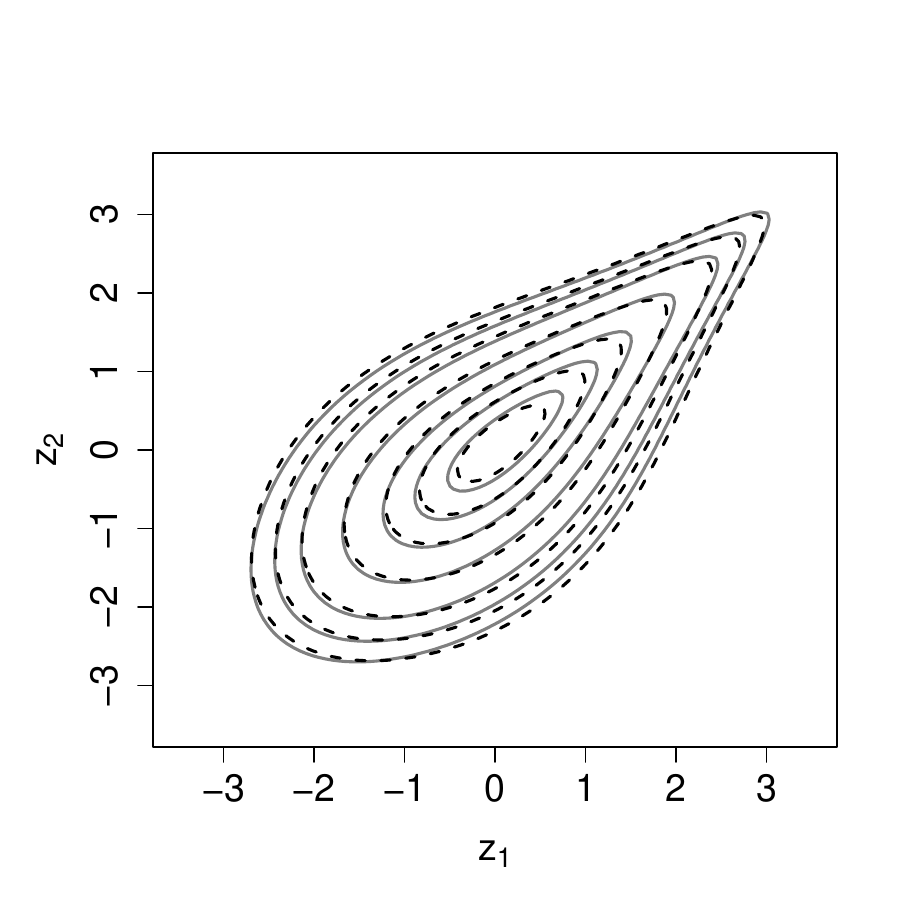}
\caption{Contour plots of the marginal density of $(\Phi^{-1}(U_1),\Phi^{-1}(U_3))^\prime$, where $U_1,...,U_4$ are distributed according to a hierarchical Kendall copula (black dashed line) or hierarchical Archimedean copula (gray line) with Gumbel cluster and nesting copulas.}
\label{fig:contarch}
\end{figure}

\begin{Example}[Bivariate margin of hierarchical Archimedean and Kendall copulas]

Let $U_1,...,U_4$ be distributed according to a hierarchical Kendall copula or hierarchical Archimedean copula with Gumbel cluster and nesting copulas: as above, $(U_1,U_2)^\prime\sim C_1$, $(U_3,U_4)^\prime\sim C_2$ and $C_0$ denotes the nesting copula.
Figure \ref{fig:contarch} illustrates the marginal density of $(U_1,U_3)^\prime$ in both cases.
It shows contour plots of the marginal density of $(\Phi^{-1}(U_1),\Phi^{-1}(U_3))^\prime$, where $\Phi^{-1}$ denotes the inverse standard normal distribution function.
Parameters are chosen as $\theta_1=3$ (Kendall's $\tau$ of $0.67$) and $\theta_2=4$ (Kendall's $\tau$ of $0.75$) for the bivariate cluster copulas $C_1$ and $C_2$, respectively, and $\theta_0=2$ (Kendall's $\tau$ of $0.5$) for the nesting copula $C_0$.
The difference between the distributions is minor.
However, the contour plot corresponding to the hierarchical Archimedean copula is slightly sharper in the upper right corner, implying a stronger joint tail behavior.

\end{Example}

Note that in the example the parameters of the cluster copulas are larger than that of the nesting copula, as required for the hierarchical Archimedean copula to yield a valid multivariate distribution ($\theta_0\leq\min\{\theta_1,\theta_2\}$).
This is however not needed for the hierarchical Kendall copula.

Finally, the nesting copula $C_0$ of a hierarchical Kendall copula is also not closed under addition and removal of cluster components $U_\ell$, which is contrary to hierarchical Archimedean copulas.
This is because the Kendall distribution function is not independent with respect to the dimension (see Expression \ref{eq:kendallrecurs}).
That is, if a random variable $U_{n+1}$ is added to cluster $i\in\{1,...,d\}$, the transformation $K_i$ and thus $V_i$ change, even if $C_i$ is Archimedean; similarly if a random variable is removed from a cluster.

%---------------------------------------------------------%

\section{Inference for hierarchical Kendall copulas}\label{sect:inference}

We now discuss appropriate statistical inference techniques for hierarchical Kendall copulas.
While hierarchical Kendall copulas could in principle be built up by any possible copula and the developed methodology holds in general, we focus in the following on the case of Archimedean clusters (see Remark \ref{rem:archclust} and Example \ref{ex:archclust}).
In this case, analytical derivations are possible due to the tractability of the Kendall distribution function \eqref{eq:archkendall} and the copula itself \eqref{eq:archcop}.
We first treat simulation, then estimation and model selection.

\subsection{Simulation}\label{sect:sim}

The following general simulation procedure describes how to sample from a given hierarchical Kendall copula with clusters that are not necessarily Archimedean.

\begin{Algorithm}[Simulation of hierarchical Kendall copulas]\label{alg:simhiercop}

Let $C_\mathcal{K}$ be a hierarchical Kendall copula with nesting copula $C_0$ and cluster copulas $C_1,...,C_d$.
Further, let $K_i^{-1}$ denotes the inverse of the Kendall distribution function $K_i$ for $i=1,...,d$.

\begin{enumerate}

\item Sample $(v_1,...,v_d)^\prime$ from $C_0$.

\item Set $z_i:=K_i^{-1}(v_i)\  \forall i\in\{1,...,d\}$.

\item Sample $(u_{m_{i-1}+1},...,u_{m_i})^\prime$ from $(U_{m_{i-1}+1},...,U_{m_i})^\prime|C_i(U_{m_{i-1}+1},...,U_{m_i})=z_i$ for $i=1,...,d$.

\item Return $\boldsymbol{u}:=(u_1,...,u_{n})^\prime$.

\end{enumerate}

\end{Algorithm}

\begin{Remark}[Simulation of $k$-level hierarchical Kendall copulas]

The above procedure for two levels can be iterated to simulate from a $k$-level hierarchical Kendall copula (see Remark \ref{def:klevelhiercop}).
In the following, samples of random variables will be denoted in the corresponding lower case letters.
Let $(v_1^{(k-1)},...,v_{d_{k-1}}^{(k-1)})^\prime$ be sampled from $C_0$.
Then,
\begin{enumerate}

\item sample from $\big(V_{m_{i-1}^{(j)}+1}^{(j-1)},...,V_{m_i^{(j)}}^{(j-1)}\big)^\prime| C_i^{(j)}\big(V_{m_{i-1}^{(j)}+1}^{(j-1)},...,V_{m_i^{(j)}}^{(j-1)}\big)=\big(K_i^{(j)}\big)^{-1}(v_i^{(j)})$ for $j=k-1,...,2$ and $i=1,...,d_j$.

\item and finally from $\big(U_{m_{i-1}^{(1)}+1},...,U_{m_i^{(1)}}\big)^\prime|C_i^{(1)}\big(U_{m_{i-1}^{(1)}+1},...,U_{m_i^{(1)}}\big)=\big(K_i^{(1)}\big)^{-1}(v_i^{(1)})$ for $i=1,...,d_1$.

\end{enumerate}

\end{Remark}

\begin{figure}
\centering
\includegraphics[width=0.325\linewidth]{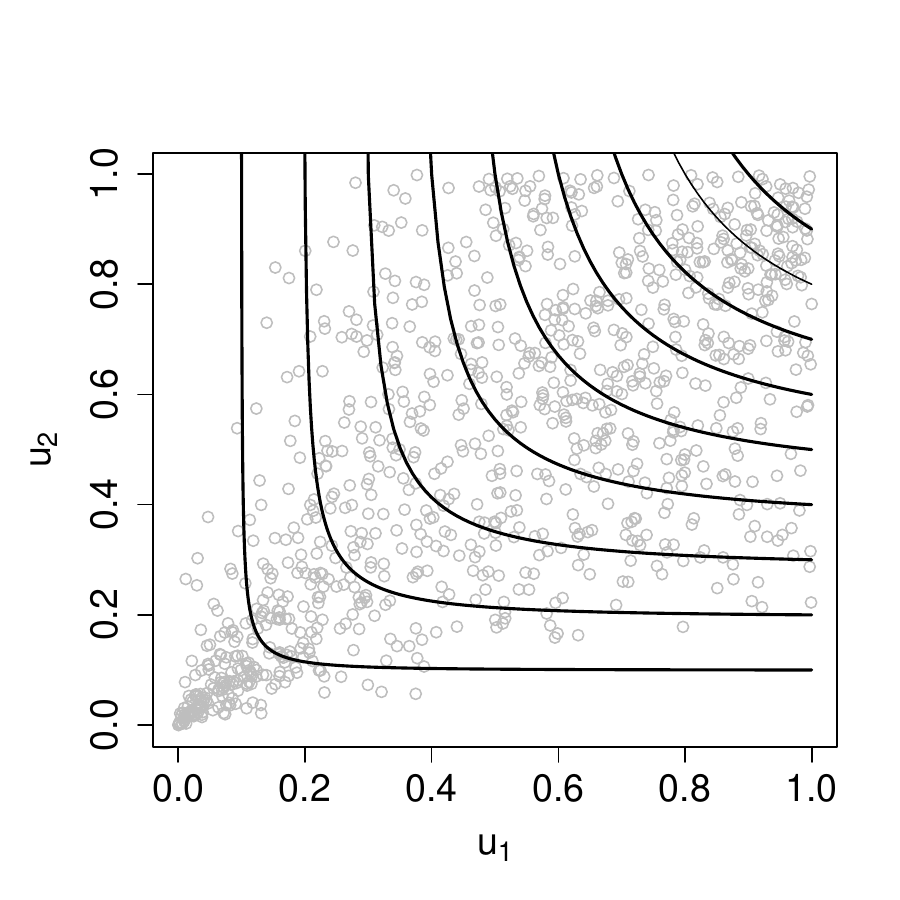}
\caption{A sample from a bivariate Clayton copula with parameter $\theta=2$ (Kendall's $\tau$ of $0.5$) and level sets $L_C(z)$ at levels $z=0.1,...,0.9$ (solid lines).}
\label{fig:level}
\end{figure}

Given that we can simulate from the copula $C_0$, sampling from hierarchical Kendall copulas thus amounts to the more general question of sampling from a distribution $\boldsymbol{U}|C(\boldsymbol{U})=z$, where $C$ is the copula of a marginally uniform random vector $\boldsymbol{U}:=(U_1,...,U_d)^\prime$ and $z\in(0,1)$.
In other words, we want to sample from a multivariate distribution given a specific level set \eqref{eq:levelset} at level $z$ as illustrated in Figure \ref{fig:level}.
This problem is discussed in the following two sections.
First, we solve the issue using the conditional inverse method, for which expressions of conditional distribution functions are provided, which are shown to be available in closed form for Archimedean copulas.
An alternative solution based on the representation given in Equation \eqref{eq:arch} is provided thereafter.
Closed-form sampling procedures for Plackett and Archimax copulas, which include the large class of extreme value copulas, can be found in \citeN{Brechmann2012}.
There, different approximate methods such as rejection sampling for the case that the cluster copulas do not admit a closed-form approach are also discussed and evaluated.

Note that inversion of the Kendall distribution function, as required in Step $(ii)$ of Algorithm \ref{alg:simhiercop}, is numerically feasible for an Archimedean copula, since its Kendall distribution function is known in closed form and can be computed efficiently (see \shortciteN{HofertMaechlerMcNeil2012} for explicit functional expressions of $(\varphi^{-1})^{(i)}$ for common Archimedean generators).

\subsubsection{Conditional inverse method}\label{sect:condinv}

A common method to generate samples from a multivariate distribution is the conditional inverse method (see \citeN{Devroye1986}).
For this, we need to determine the iterative conditional distribution functions of $\boldsymbol{U}|C(\boldsymbol{U})=z$, that is, of $U_1|C(\boldsymbol{U})=z$, $U_2|(U_1=u_1, C(\boldsymbol{U})= z)$, ..., $U_{d-1}|(U_1=u_1, ..., U_{d-2}=u_{d-2}, C(\boldsymbol{U})= z)$.
%\footnote{The distribution of $U_{d}|(U_1=u_1, ..., U_{d-1}=u_{d-1}, C(\boldsymbol{U})= z)$ actually does not need to be determined because the value of $U_d$ is uniquely given through the conditioning variables.}
The corresponding conditional distribution functions are denoted by $F_{j|1,...,j-1}(\cdot|u_1,...,u_{j-1},z)$ and densities by $f_{j|1,...,j-1}(\cdot|u_1,...,u_{j-1},z)$ for $j=1,...,d-1$, respectively.
Then we obtain observations $(u_1,...,u_d)^\prime$ from $\boldsymbol{U}|C(\boldsymbol{U})= z$ by generating $w_1,...,w_{d-1}$ independently from the uniform distribution and setting $u_j:=F_{j|1,...,j-1}^{-1}(w_j|u_1,...,u_{j-1},z)$ for $j=1,...,d-1$. 
For $j=d$ it is $u_d:=C_{u_1,...,u_{d-1}}^{-1}(z)$.
The problem therefore is to determine the conditional distribution functions, which are generally not given in closed form.

\begin{Theorem}[Conditional distributions]\label{theo:conddist}

Let $\boldsymbol{U}\sim C$, then for all $j=1,...,d-1,$
\begin{equation}
F_{j|1,...,j-1}(u|u_1,...,u_{j-1},z) = \dfrac{\int_{C_{u_1,...,u_{j-1}}^{-1}(z)}^u g_j(u_j)\, du_j}{\int_{C_{u_1,...,u_{j-1}}^{-1}(z)}^1 g_j(u_j)\, du_j},\quad u\in(C_{u_1,...,u_{j-1}}^{-1}(z),1),
\label{eq:intconddist}
\end{equation}
where
\begin{equation}
g_j(u_j) = \int_{C_{u_1,...,u_j}^{-1}(z)}^{1} ... \int_{C_{u_1,...,u_{d-2}}^{-1}(z)}^{1} c(u_1,...,u_{d-1},C_{u_1,...,u_{d-1}}^{-1}(z)) \frac{\partial}{\partial z} C_{u_1,...,u_{d-1}}^{-1}(z)\, du_{d-1} ... du_{j+1}.
\label{eq:intconddens}
\end{equation}

\end{Theorem}

\noindent
\textbf{Proof:} The idea is to derive the conditional density $f_{j|1,...,j-1}(\cdot|u_1,...,u_{j-1},z)$ and then integrate to obtain the distribution function.
We begin by observing that
\begin{equation}
f_{j|1,...,j-1}(u_j|u_1,...,u_{j-1},z) = \frac{f_{U_1,...,U_j,C(\boldsymbol{U})}(u_1,...,u_{j-1},u_j,z)}{f_{U_1,...,U_{j-1},C(\boldsymbol{U})}(u_1,...,u_{j-1},z)}.
\label{eq:intconddensdecomp}
\end{equation}
According to a change of variables as in Equation \eqref{eq:jointdens1b}, the numerator can then be rewritten as
\begin{align*}
& f_{U_1,...,U_j,C(\boldsymbol{U})}(u_1,...,u_j,z)\\
& = \int_{C_{u_1,...,u_j}^{-1}(z)}^{1} ... \int_{C_{u_1,...,u_{d-2}}^{-1}(z)}^{1} f_{U_1,...,U_{d-1},C(\boldsymbol{U})}(u_1,...,u_{d-1},z)\, du_{d-1} ... du_{j+1}\\
& = \int_{C_{u_1,...,u_j}^{-1}(z)}^{1} ... \int_{C_{u_1,...,u_{d-2}}^{-1}(z)}^{1} c(u_1,...,u_{d-1},C_{u_1,...,u_{d-1}}^{-1}(z)) \frac{\partial}{\partial z} C_{u_1,...,u_{d-1}}^{-1}(z)\, du_{d-1} ... du_{j+1}\\
& = g_j(u_j),
\end{align*}
where $g_j$ is defined in Equation \eqref{eq:intconddens} and dependence on $u_1,...,u_{j-1},z$ is suppressed for ease of notation.
Further, the denominator of Equation \eqref{eq:intconddensdecomp} then reads as
\begin{equation*}
f_{U_1,...,U_{j-1},C(\boldsymbol{U})}(u_1,...,u_{j-1},z) = \int_{C_{u_1,...,u_{j-1}}^{-1}(z)}^1 g_j(u_j)\, du_j.
\end{equation*}
By integration, we obtain the expression for the conditional distribution function \eqref{eq:intconddist}.\hfill$\square$\\

Evidently, the conditional distribution functions given in Equation \eqref{eq:intconddist} in general do not allow for explicit expressions.
In particular, if the copula quantile function $C^{-1}$ is not available in closed form such as for the Gaussian copula, the expression in Equation \eqref{eq:intconddist} hardly simplifies.
In the case of Archimedean copulas, the conditional distribution functions can be obtained as particularly convenient expressions.

\begin{Lemma}[Conditional distributions of Archimedean copulas]\label{lem:archconddist}

Let $\boldsymbol{U}\sim C$, where $C$ is a $d$-dimensional Archimedean copula with generator $\varphi$, then for all $j=1,...,d-1,$
\begin{equation}
F_{j|1,...,j-1}(u|u_1,...,u_{j-1},z) = \left(1-\frac{\varphi(u)}{\varphi(z)-\sum_{1 \leq i < j} \varphi(u_i)} \right)^{d-j},\quad u\in(C_{u_1,...,u_{j-1}}^{-1}(z),1).
\label{eq:archconddist}
\end{equation}

\end{Lemma}

This result can be derived using the general formula provided in Theorem \ref{theo:conddist}.
As noted by an anonymous referee, a considerably simpler proof can be formulated by exploiting the representation \eqref{eq:arch} of Archimedean copulas and properties of the Dirichlet distribution.
The proof is given in the next section in Remark \ref{rem:archconddist} using a characterization result of Proposition \ref{prop:projectedarch} below.

Lemma \ref{lem:archconddist} then allows to use the conditional inverse method for Archimedean copulas, for which the conditional distribution functions can easily be inverted in closed form.

\begin{Algorithm}[Conditional inverse method for Archimedean copulas]\label{alg:archcondinv}

Let $C$ be an Archimedean copula with generator $\varphi$ and $z\in(0,1)$.

\begin{enumerate}

\item Sample $w_1,...,w_{d-1}$ independently from the uniform distribution.

\item For $j=1,...,d-1$: $u_j:=\varphi^{-1}((1-w_j^{1/(d-j)})(\varphi(z)-\sum_{1 \leq i < j} \varphi(u_i)))$.

\item Set $u_d:=\varphi^{-1}(\varphi(z)-\sum_{1 \leq i < d} \varphi(u_i))$.

\item Return $(u_1,...,u_d)^\prime$.

\end{enumerate}

\end{Algorithm}

\begin{figure}[t]
\centering
\includegraphics[width=0.325\linewidth]{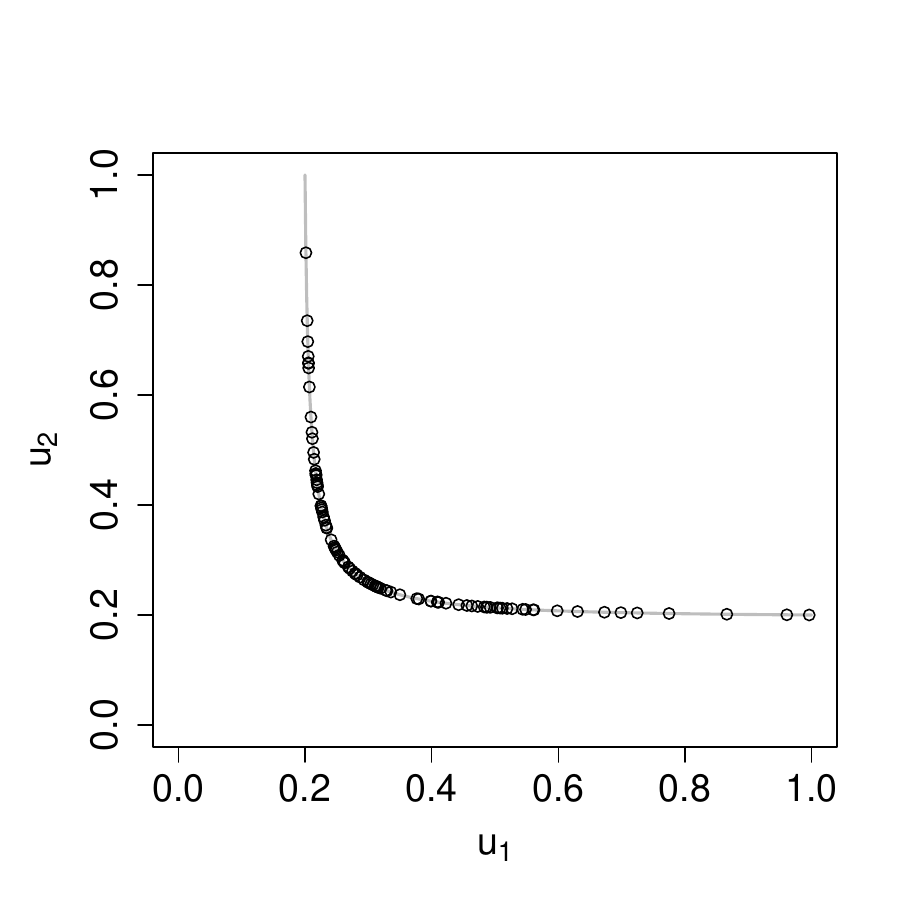}
\includegraphics[width=0.325\linewidth]{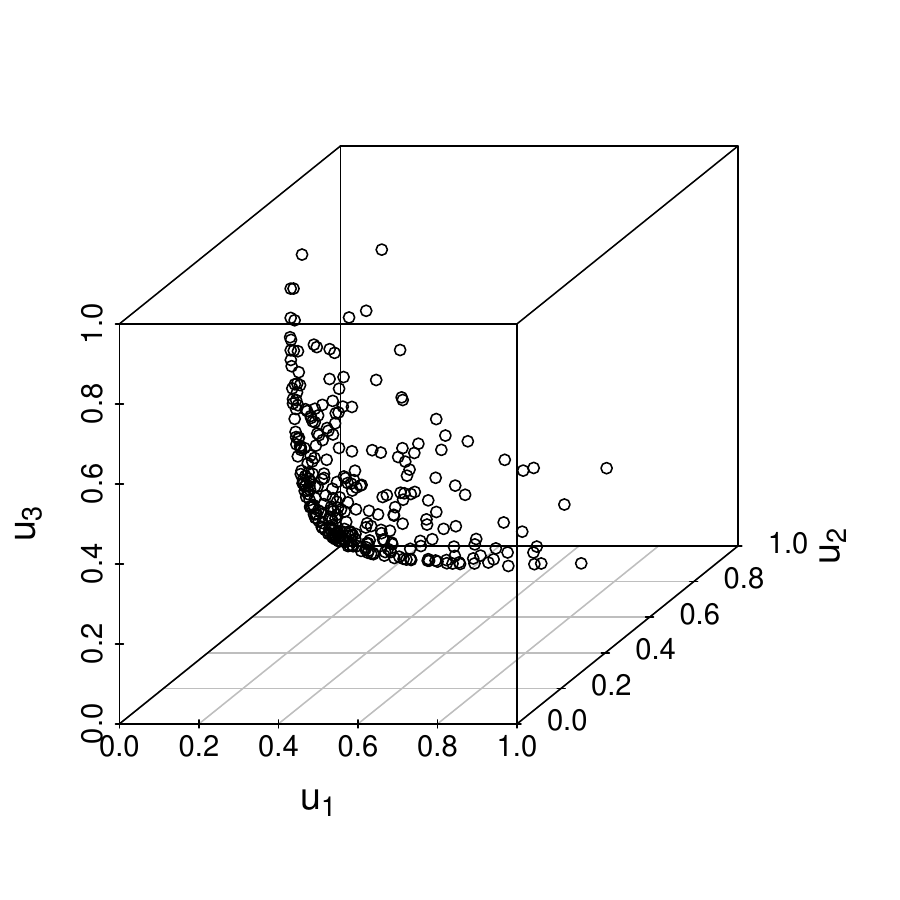}
\includegraphics[width=0.325\linewidth]{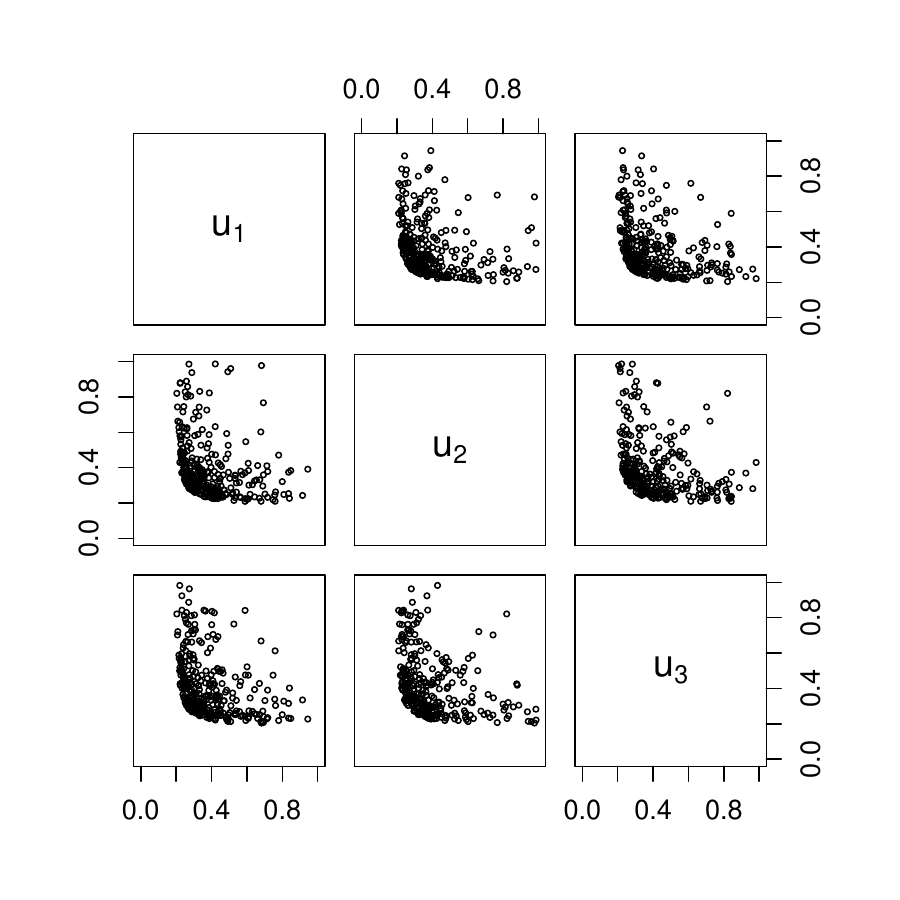}
\caption{Left panel: scatter plot of a sample of a bivariate Clayton copula with parameter $\theta=2$ at $z=0.2$. Middle and right panel: 3D scatter plot and pairwise scatter plots of a sample of a trivariate Clayton copula with parameter $\theta=2$ at $z=0.2$.}
\label{fig:archsim}
\end{figure}

For illustration Figure \ref{fig:archsim} shows scatter plots of samples from bivariate and trivariate Clayton copulas with parameter $\theta=2$ (Kendall's $\tau$ of $0.5$).

As a side note, we observe that Algorithm \ref{alg:archcondinv} can in particular be used to sample from a given Archimedean copula.
For this, draw an additional uniform observation $w_d$ and set $z:=K^{-1}(w_d)$ prior to preforming Steps $(ii)$ and $(iii)$.
An equivalent version of this result has previously been stated in \shortciteN{WuValdezSherris2007}.

\subsubsection{Projected distribution}

Clearly the condition $C(\boldsymbol{U})= z$ on the distribution of $\boldsymbol{U}$ means that we are in fact investigating a $(d-1)$-dimensional distribution, namely the distribution of $\boldsymbol{U}$ projected to the $(d-1)$-dimensional level set $L_C(z)\subset[0,1]^d$.
This distribution is however not easily tractable in general.

Again in the case of Archimedean copulas some convenient results are obtainable.

\begin{Proposition}[Projected distribution of Archimedean copulas]\label{prop:projectedarch}

Let $\boldsymbol{U}\sim C$, where $C$ is a $d$-dimensional Archimedean copula with generator $\varphi$, then it holds for $z\in(0,1)$ that
\begin{equation}
[\boldsymbol{U}|C(\boldsymbol{U})= z] \stackrel{d}{=} \left(\varphi^{-1}(S_1\varphi(z)),...,\varphi^{-1}(S_d\varphi(z))\right)^\prime,
\label{eq:projectedarch}
\end{equation}
where $\boldsymbol{S}=(S_1,...,S_d)^\prime$ is uniformly distributed on the unit simplex.

\end{Proposition}

\noindent
\textbf{Proof:} According to Equation \eqref{eq:arch}, we have the representation $(\varphi(U_1),...,\varphi(U_d))^\prime \stackrel{d}{=} R\boldsymbol{S}$, where $R=\sum_{j=1}^d \varphi(U_j)=\varphi(C(\boldsymbol{U}))$ is the radial part, which is independent of $\boldsymbol{S}$.
Fixing the level set $L_C(z)$ is therefore equivalent to setting $R=\varphi(z)$, so that we obtain Equation \eqref{eq:projectedarch}. \hfill$\square$\\

Equation \eqref{eq:projectedarch} for the projected distribution of an Archimedean copula is particularly appealing, since it does not depend on the radial variable and its distribution, which may not be available in closed form.
This result can then be used to provide another sampling algorithm for $\boldsymbol{U}|C(\boldsymbol{U})=z$, which can be shown to be equivalent to Algorithm \ref{alg:archcondinv}, when using explicit expressions for the observations $(s_1,...,s_d)^\prime$ from $\boldsymbol{S}$ in terms of uniform random variables (see \citeN[Lemma 3.1.8]{Hering2011}).

\begin{Algorithm}[Projected distribution sampling for Archimedean copulas]\label{alg:projectedsim}

Let $C$ be an Archimedean copula with generator $\varphi$ and $z\in(0,1)$.

\begin{enumerate}

\item Sample $(s_1,...,s_d)^\prime$ from $\boldsymbol{S}$.

\item For $j=1,...,d$: $u_j:=\varphi^{-1}(s_j\varphi(z))$.

\item Return $(u_1,...,u_d)^\prime$.

\end{enumerate}

\end{Algorithm}

%When also sampling the contour level $z$ (for instance by conditional inverse sampling using the Kendall distribution function as mentioned above) to obtain observations from the Archimedean copula, this is again the approach proposed by \shortciteN{WuValdezSherris2007} which has been restated by \citeN{Hering2011} in the setting of the work by \citeN{McNeilNeslehova2009} as we use it here.

As stated above, it is an open problem to determine this projected distribution for general copulas.
Analogous sampling methods to Algorithm \ref{alg:projectedsim} could be used then.

Finally, Proposition \ref{prop:projectedarch} also enables us to formulate a simple proof of Lemma \ref{lem:archconddist}.

\begin{Remark}[Proof of Lemma \ref{lem:archconddist}]\label{rem:archconddist}

We show that it holds for an Archimedean copula $C$ and all $j=1,...,d-1$ that
\begin{equation*}
F_{j|1,...,j-1}(u|u_1,...,u_{j-1},z) = \left(1-\frac{\varphi(u)}{\varphi(z)-\sum_{1 \leq i < j} \varphi(u_i)} \right)^{d-j},\quad u\in(C_{u_1,...,u_{j-1}}^{-1}(z),1).
\end{equation*}
According to Proposition \ref{prop:projectedarch}, it holds that
\begin{align*}
F_{j|1,...,j-1}(u|u_1,...,u_{j-1},z) & = P(U_j\leq u|U_1=u_1,...,U_{j-1}=u_{j-1},C(\boldsymbol{U})=z)\\
& = P\left(S_j\geq \left.\frac{\varphi(u)}{\varphi(z)}\right|S_1=\frac{\varphi(u_1)}{\varphi(z)},...,S_{j-1}=\frac{\varphi(u_{j-1})}{\varphi(z)}\right),
\end{align*}
where $\boldsymbol{S}=(S_1,...,S_d)^\prime$ follows a uniform distribution on the unit simplex, which can be represented as a Dirichlet distribution with all parameters equal to 1.
It then holds that (see \shortciteN[Theorem 1.6]{FangKotzNg1990})
\begin{equation*}
\left.\frac{S_j}{1-s_1-...-s_{j-1}}\right|\left(S_1=s_1,...,S_{j-1}=s_{j-1}\right) \sim \text{Beta}(1,d-j),\quad j=1,...,d-1.
\end{equation*}
Further, the distribution function of the $\text{Beta}(1,d-j)$ distribution is $F_{\text{Beta}}(s;1,d-j)=1-(1-s)^{d-j}$.
Therefore, we obtain
\begin{align*}
P&\left(S_j\geq \left.\frac{\varphi(u)}{\varphi(z)}\right|S_1=\frac{\varphi(u_1)}{\varphi(z)},...,S_{j-1}=\frac{\varphi(u_{j-1})}{\varphi(z)}\right)\\
& = 1-F_{\text{Beta}}\left(\frac{\frac{\varphi(u)}{\varphi(z)}}{1-\frac{\varphi(u_1)}{\varphi(z)}-....-\frac{\varphi(u_{j-1})}{\varphi(z)}};1,d-j\right)\\
& = \left(1-\frac{\varphi(u)}{\varphi(z)-\sum_{1 \leq i < j} \varphi(u_i)} \right)^{d-j},
\end{align*}
as claimed in Lemma \ref{lem:archconddist}.

\end{Remark}

\subsection{Estimation}\label{sect:est}

In light of Sklar's theorem \eqref{eq:sklar}, it is common in dependence modeling to transform data $(x_{j1},...,x_{jn})^\prime,\ j=1,...,N,$ to $[0,1]^n$ using the marginal distribution functions $F_{X_i},\ i=1,...,n$, that is, we compute $u_{ji}=F_{X_i}(x_{ji})$.
In most cases, $F_{X_i}$ will however be unknown so that this transformation needs to be based on a parametric or non-parametric estimate, which introduces uncertainty into the modeling.

When a parametric modeling of the margins is chosen, parameters of margins and dependence model can either be estimated jointly or, when this is not feasible, sequentially using the estimation method of inference functions for margins (IFM) by \citeN{McLeishSmall1988} and \citeN{JoeXu1996}.
In the IFM method, first marginal parameters are estimated and then dependence parameters given the estimated margins $\widehat{F}_{X_i},\ i=1,...,n$.
%That is, if we set $\widehat{u}_{ji}=\widehat{F}_i(x_{ji})$, then we estimate the parameters $\boldsymbol{\theta}_0,\boldsymbol{\theta}_1,...,\boldsymbol{\theta}_d$ of a hierarchical Kendall copula $C_{\mathcal{K}}$ with nesting and cluster copulas $C_0,C_1,...,C_d$ and joint density function $c_{\mathcal{K}}$ \eqref{eq:jointdens} (under Assumptions $\mathcal{A}_1$ and $\mathcal{A}_2$) by maximizing the log likelihood expression
%\begin{equation}
%\begin{split}
%\ell_{\mathcal{K}}&(\boldsymbol{\theta}_0,\boldsymbol{\theta}_1,...,\boldsymbol{\theta}_d) = \sum_{j=1}^N \log c_{\mathcal{K}}(\widehat{u}_{j1},...,\widehat{u}_{jn}|\boldsymbol{\theta}_0,\boldsymbol{\theta}_1,...,\boldsymbol{\theta}_d)\\
%& = \sum_{j=1}^N \log c_0(K_1(C_1(\widehat{u}_{j1},...,\widehat{u}_{jm_1}|\boldsymbol{\theta}_1)|\boldsymbol{\theta}_1),...,K_d(C_d(\widehat{u}_{jm_{d-1}+1},...,\widehat{u}_{jn}|\boldsymbol{\theta}_d)|\boldsymbol{\theta}_d)|\boldsymbol{\theta}_0)\\
%&\quad + \sum_{j=1}^N \sum_{i=1}^d \log c_i(\widehat{u}_{jm_{i-1}+1},...,\widehat{u}_{jm_i}|\boldsymbol{\theta}_i)\\
%&=: \ell_0(\boldsymbol{\theta}_0,\boldsymbol{\theta}_1,...,\boldsymbol{\theta}_d) + \sum_{i=1}^d \ell_i(\boldsymbol{\theta}_i),
%\end{split}
%\label{eq:jointlik}
%\end{equation}
That is, parameters of a hierarchical Kendall copula are estimated based on $\widehat{u}_{ji}=\widehat{F}_{X_i}(x_{ji}),\ j=1,...,N,\ i=1,...,n,$ by maximizing the log likelihood, which conveniently decomposes into separate sums (see Theorem \ref{theo:joint}) and which is straightforward to evaluate for Archimedean cluster copulas with Kendall distribution function given in Equation \eqref{eq:archkendall}.

The asymptotic covariance matrix is given by the inverse Godambe information matrix, which unfortunately is typically very cumbersome to compute.
To see this, note that the joint density \eqref{eq:jointdens} depends on the parameters of a cluster copula both through the density of the cluster copula as well as through the arguments of the nesting copula.
For such situations, \citeN{JoeXu1996} propose a jackknife estimate of the asymptotic covariance.
In our application in Section \ref{sect:applic}, the margins will be time-dependent.
In this case, a stationary block bootstrap can be used to calculate approximate standard errors (see \citeN{PolitisRomano1994} and \citeN{GoncalvesWhite2004}).

As an approximation to maximum likelihood estimates of the dependence parameters we additionally propose a sequential approach and evaluate and compare the finite sample behavior of the estimators in an extensive simulation study.

\subsubsection{Sequential estimation}

The hierarchical construction given in Definition \ref{def:hiercop}
%and the log likelihood expression \eqref{eq:jointlik}
directly leads to a sequential estimation procedure of hierarchical Kendall copulas, which avoids higher dimensional maximum likelihood estimation.

\begin{Algorithm}[Sequential estimation of hierarchical Kendall copulas]\label{alg:seqest}

Let $(u_{j1},...,u_{jn})^\prime_{j=1,...,N}$ be a sample of a hierarchical Kendall copula $C_\mathcal{K}$ as defined in Definition \ref{def:hiercop} (possibly after appropriate transformation of the margins).
Further, let $\boldsymbol{\theta}_0,\boldsymbol{\theta}_1,...,\boldsymbol{\theta}_d$ be the parameters of the copulas $C_0,C_1,...,C_d$, respectively.
We then obtain corresponding sequential estimates $\widehat{\boldsymbol{\theta}}_i,\ i=0,...,d,$ as follows.

\begin{enumerate}

\item For each $i\in\{1,...,d\}$ estimate $\boldsymbol{\theta}_i$ based on $(u_{j,m_{i-1}+1},...,u_{j,m_i})^\prime_{j=1,...,N}$ by maximum likelihood.
%, that is maximize $\ell_i(\boldsymbol{\theta}_i)$ as defined in \eqref{eq:jointlik} with respect to $\boldsymbol{\theta}_i$.

\item Estimate $\boldsymbol{\theta}_0$ based on the pseudo observations
\begin{equation} \widehat{v}_{ji}:=K_i(C_i(u_{j,m_{i-1}+1},...,u_{j,m_i};\widehat{\boldsymbol{\theta}}_i);\widehat{\boldsymbol{\theta}}_i),\ i=1,...,d,\ j=1,...,N,
\label{eq:pseudoobs}
\end{equation}
by maximum likelihood.
%, that is maximize $\ell_0(\boldsymbol{\theta}_0,\widehat{\boldsymbol{\theta}}_1,...,\widehat{\boldsymbol{\theta}}_d)$ as defined in \eqref{eq:jointlik} with respect to $\boldsymbol{\theta}_i$.

\end{enumerate}

\end{Algorithm}

Clearly, this two-step estimation procedure directly generalizes to a $k$-step estimation approach for $k$-level hierarchical Kendall copulas (see Remark \ref{def:klevelhiercop}).
Resulting estimates may be used as starting values for a joint maximum likelihood estimation of the dependence parameters.

\subsubsection{Simulation study}\label{sect:estsimstudy}

\begin{figure}
\centering
\includegraphics[width=.9\linewidth]{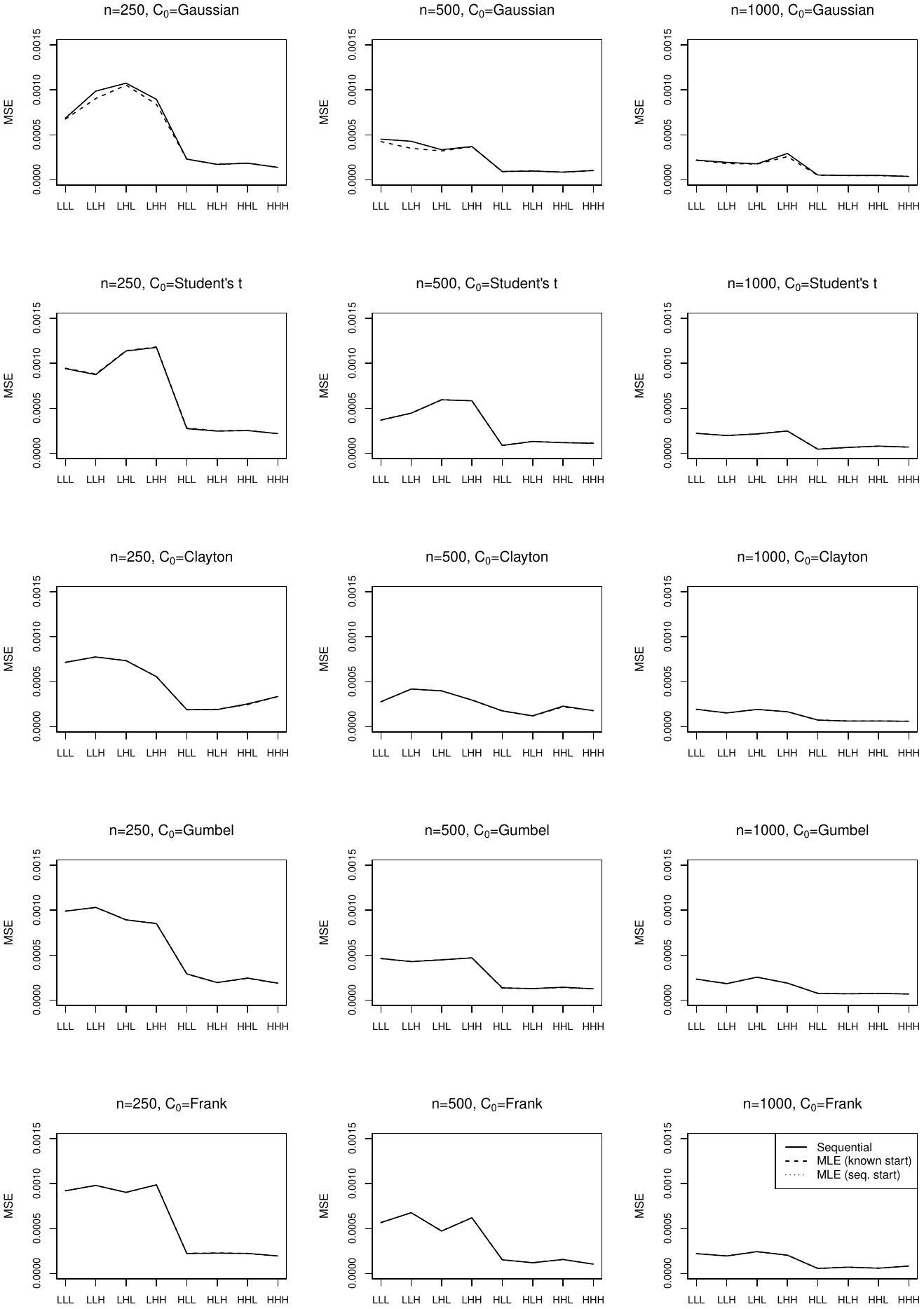}
\caption{Mean squared errors (MSEs) of estimated parameters in terms of Kendall's $\tau$ values of the three estimation procedures for different sample sizes $n$ and nesting copulas $C_0$.
Cluster copula 1: Clayton. Cluster copula 2: Gumbel.
Notation for the $x$-axis: ($\tau_0$,$\tau_1$,$\tau_2$), where $L:=0.4$ and $H:=0.7$. The range of the $y$-axes is chosen such that MSEs in each column are comparable.}
\label{fig:simest2}
\end{figure}

In order to examine the finite sample behavior of the estimation procedures discussed above, we perform a large scale Monte Carlo study.
For this, we simulate from a four-dimensional hierarchical Kendall copula (two bivariate clusters; margins are assumed to be known) and then estimate the parameters according to the following methods:
\begin{itemize}

\item Sequential estimation;

\item MLE with known starting values (true parameters);

\item MLE with sequentially estimated starting values.

\end{itemize}
The cluster copulas $C_1$ and $C_2$ are chosen as Clayton, Gumbel or Frank; the nesting copula $C_0$ as Gaussian, Student's t (ten degrees of freedom), Clayton, Gumbel or Frank.
Parameters are determined according to Kendall's $\tau$ values of $0.4$ and $0.7$.
Sample sizes are 250, 500 and 1000 and the number of repetitions is 100.
Estimation accuracy is compared based on the mean squared error of the estimated nesting copula parameter $\theta_0$ (transformed to Kendall's $\tau$ values)
as shown in Figure \ref{fig:simest2} for the case of Clayton and Gumbel cluster copulas.
Results of the other cases are available from the author upon request.
An illustrative sample of size 1000 for the case of Clayton and Gumbel cluster copulas (Kendall's $\tau$ of $0.4$) and Frank nesting copula (Kendall's $\tau$ of $0.7$) is shown in Figure \ref{fig:exarch}.

It turns out that results are essentially independent of the chosen cluster copulas (not shown here).
Similarly, the choice of parameters of the cluster copulas does not have a great influence on the results, while larger nesting parameters mean more accurate results.
Overall, there is hardly any difference between the three estimation procedures.
In particular, this means that sequential estimation provides good starting values for joint estimation of the dependence parameters.

\subsection{Model selection}\label{sect:modselect}

In practical applications, the clusters $(U_{m_{i-1}+1},...,U_{m_i})^\prime,\ i\in\{1,...,d\},$ have to be identified.
In cases where they are not given from the data (e.g., industry sectors in financial data; see Section \ref{sect:applic}), common clustering techniques (see, e.g., \shortciteN{HastieTibshiraniFriedman2009}) can be used.
If a multi-level hierarchical Kendall copula is considered, hierarchical clustering methods may be particularly helpful.
In hierarchical clustering, the use of an appropriate metric to measure the closeness between (groups of) variables is essential.
Inspired by \citeN{Mantegna1999}, who uses the linear correlation coefficient,  we propose to use the following metric between variables $j$ and $k$,
\begin{equation}
d(j,k) = \sqrt{1-\widehat{\rho}_{jk}^S},
\label{eq:metric}
\end{equation}
where $\widehat{\rho}_{jk}^S$ is the empirical Spearman's $\rho$ between the variables.
Thus, the stronger the dependence between variables $j$ and $k$, the smaller is $d(j,k)$.
Obviously, $d(j,k)=0$ if variables $j$ and $k$ are comonotonic, that is, if the observations are the same.
Further, it holds that $d(j,k)=d(k,j)$ (symmetry) and $d(j,k)\leq d(j,\ell) + d(\ell,k)$ for another variable $\ell$.

Hierarchical clustering further requires the choice of a linkage criterion to determine the distance between groups of variables.
Classical average linkage clustering simply uses the mean distance between the elements of the groups.
In the setting of hierarchical Kendall copulas, it is however more natural to form the pseudo observations $\widehat{v}_{ji}$ \eqref{eq:pseudoobs} based on purely empirical versions of the copula $C_i$ and the Kendall distribution function $K_i$ (see, e.g., \shortciteN{BarbeGenestGhoudiRemillard1996}) and then compute the distance \eqref{eq:metric} between them, since this respects the model definition.
This measure of association between multivariate random vectors, along with others, is also discussed by \shortciteN{GrotheSchmidSchniedersSegers2011}.
It however does not ensure that the closeness between grouped variables is monotone decreasing with increasing level of the merger.
The process of the merger is typically illustrated in a binary tree, which is called dendogram and represents the closeness between cluster members (see Figure \ref{fig:cluster} below).

Recalling the discussion in Example \ref{ex:kendall}, the size of clusters has to be carefully chosen because Kendall distributions may become almost degenerate at $0$ for very large clusters.
In most practical situations, this is however not an issue, since already under medium positive dependence the convergence to the constant function at 1 is very slow (see the right panel of Figure \ref{fig:Kindgumbel}).

In the next step, copulas have to be selected for the clusters.
Due to the hierarchical nature of the model, higher order levels depend on copulas in lower levels, so that a careful selection of cluster copulas is necessary.
A possible approach is a stepwise selection similar to the sequential estimation procedure outlined in Algorithm \ref{alg:seqest}, that is, the nesting copula $C_0$ is selected based on the pseudo observations \eqref{eq:pseudoobs}.
This is similar to selection approaches of hierarchical Archimedean copulas \shortcite{OkhrinOkhrinSchmid2012} and of pair copula constructions \shortcite{DissmannBrechmannCzadoKurowicka2013}.
Common criteria for the selection of copulas are information criteria such as the AIC.
%Goodness-of-fit tests should then be used to verify the fit (see \citeN{Berg2009} and \shortciteN{GenestRemillardBeaudoin2009} for two overview studies).
Since the selection based on pseudo observations however induces uncertainty in the selection of the nesting copula, we perform a misspecification study.

\subsubsection{Copula misspecification}\label{sect:misspec}

To analyze the effect of misspecification of the cluster and nesting copulas, we resume the setting of Section \ref{sect:estsimstudy} and simulate samples of size 1000 from four-dimensional hierarchical Kendall copulas with cluster copulas $C_1$ and $C_2$ chosen as Clayton, Gumbel or Frank, and the nesting copula $C_0$ as Gaussian, Student's t (ten degrees of freedom), Clayton, Gumbel or Frank.
Parameters are again determined according to Kendall's $\tau$ values of $0.4$ and $0.7$ and the number of repetitions is 100.

In addition, we simulate from a range of alternative multivariate copulas to investigate how well these copulas can be approximated by a hierarchical Kendall copula.
We consider four-dimensional regular vine pair copula constructions (see \shortciteN{DissmannBrechmannCzadoKurowicka2013} and the R package \texttt{VineCopula} by \shortciteN{VineCopula}) with unconditional pair copulas chosen as $C_{1,2}=C_1$, $C_{2,3}=C_0$ and $C_{3,4}=C_2$ and conditional pair copulas chosen as $C_{1,3|2}=C_{2,4|3}=C_{1,4|2,3}=C_0$, whose parameters are determined according to a decreasing value of Kendall's $\tau$ compared to the unconditional copula $C_{2,3}$\footnote{If $\tau_{j,k|D}$ denotes the Kendall's $\tau$ corresponding to the pair copula $C_{j,k|D}$ and $\tau_0$ is the Kendall's $\tau$ of $C_{2,3}=C_0$, then we choose $\tau_{1,3|2}=\tau_{2,4|3}=2\tau_0/3$ and $\tau_{1,4|2,3}=\tau_0/3$}.
Such a pair copula construction mimics, to some extent, a hierarchical dependence model, but without having explicit between-cluster dependence.
Furthermore, we simulate from hierarchical Archimedean copulas with Clayton, Gumbel and Frank cluster and nesting copulas.
This is however only possible if $\tau_0\leq\min\{\tau_1,\tau_2\}$, since the between-cluster dependence cannot be stronger than the within-cluster dependence.
Finally, we also consider four-dimensional Gaussian and Student's t copulas (ten degrees of freedom) with correlation matrices clustered according to the respective Kendall's $\tau$ values for within- and between-cluster dependence\footnote{In order to ensure positive definiteness, Kendall's $\tau$ values have to be adapted when between-cluster dependence is $0.7$.
Then, within-cluster dependence is set to either $0.6$ for both clusters or to $0.7$ and $0.5$ for the different clusters.}.

\begin{figure}
\centering
\includegraphics[width=.88\linewidth]{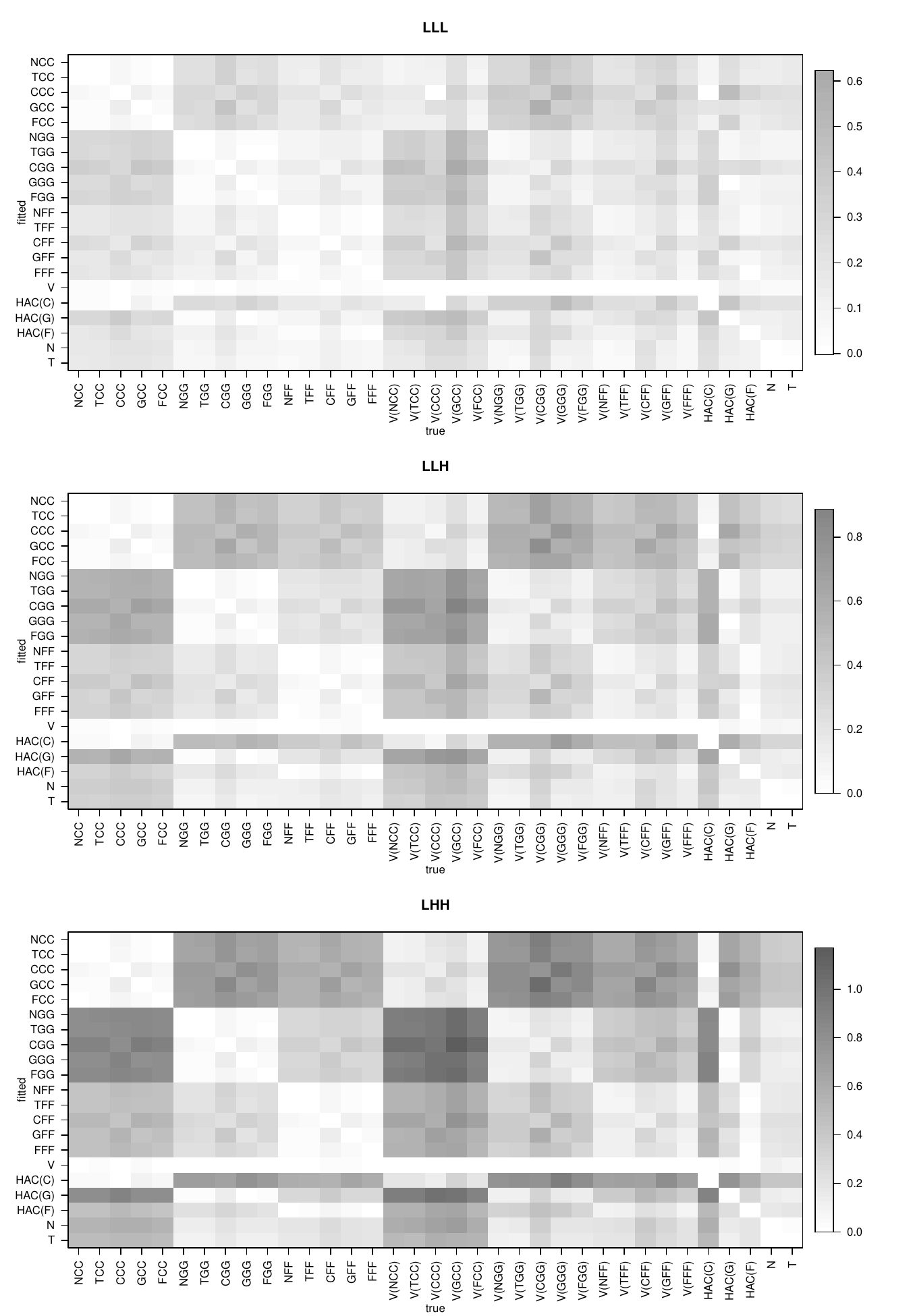}
\caption{Illustration of mean Kullback-Leibler divergences (part 1): light colors indicate a small divergence, dark colors a large divergence.
% (black corresponds to the maximum divergence of 1.839).
Notation for $x$- and $y$-axes: ($C_0$,$C_1$,$C_2$) with Gaussian (N), Student's t (T), Clayton (C), Gumbel (G), and Frank (F) copulas; vine copulas are indicated by `V', hierarchical Archimedean copulas by `HAC'.
Columns correspond to the true (simulated) models, rows to the fitted models
(only the best fitting among the 15 considered vine copulas is shown).
Notation for panel titles: ($\tau_0$,$\tau_1$,$\tau_2$), where $L:=0.4$ and $H:=0.7$.}
\label{fig:misspec}
\end{figure}

\begin{figure}
\centering
\includegraphics[width=.88\linewidth]{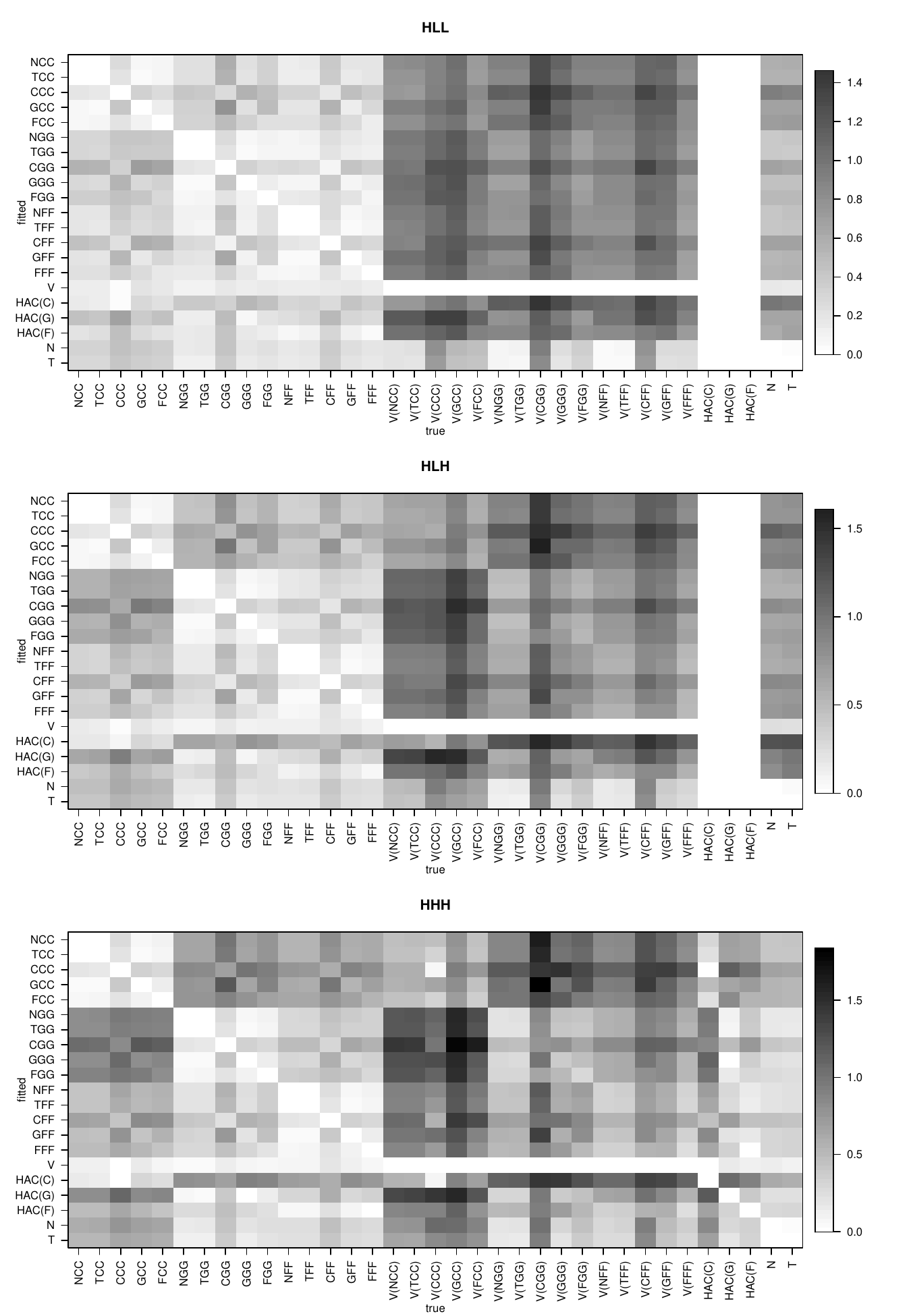}
\caption{Illustration of mean Kullback-Leibler divergences (part 2): light colors indicate a small divergence, dark colors a large divergence.
% (black corresponds to the maximum divergence of 1.839).
Notation for $x$- and $y$-axes: ($C_0$,$C_1$,$C_2$) with Gaussian (N), Student's t (T), Clayton (C), Gumbel (G), and Frank (F) copulas; vine copulas are indicated by `V', hierarchical Archimedean copulas by `HAC'.
Columns correspond to the true (simulated) models, rows to the fitted models
(only the best fitting among the 15 considered vine copulas is shown).
Notation for panel titles: ($\tau_0$,$\tau_1$,$\tau_2$), where $L:=0.4$ and $H:=0.7$.}
\label{fig:misspec2}
\end{figure}

The effect of misspecification is then examined in terms of the Kullback-Leibler divergence between the true (simulated) model and the alternative models fitted by maximum likelihood.
The results are illustrated in Figures \ref{fig:misspec} and \ref{fig:misspec2}.
First of all, they show that the hierarchical Kendall copulas provide a very good fit if the cluster copulas are identified correctly.
This means that the effect of the uncertainty with respect to the selection of the nesting copula based on pseudo observations is not severe.
These results also hold when the true model is a hierarchical Archimedean copula, which apparently can be well approximated by a hierarchical Kendall copula (see Figure \ref{fig:contarch}).
Clearly, differences between the models become more distinct with increasing dependence in terms of Kendall's $\tau$.
In particular, the Clayton copula is rather different from the other copulas (see the shape of the scatter plots in Figure \ref{fig:exarch}) and therefore harder to approximate by a misspecified model.
As a result, the elliptical copulas are also best approximated by hierarchical Kendall copulas with Gumbel and Frank components.
In particular, hierarchical Kendall copulas with elliptical nesting copula and Gumbel or Frank cluster copulas are very close in terms of the Kullback-Leibler divergence even if clusters are heterogeneous.
The non-hierarchical pair copula constructions are naturally more difficult to approximate by hierarchical Kendall copulas.
Especially when the cluster copulas $C_1$ and $C_2$ are selected in the same way as the pair copulas $C_{1,2}$ and $C_{3,4}$, respectively, hierarchical Kendall copulas may however be quite close in terms of the Kullback-Leibler divergence.
Conversely, a pair copula construction may also quite well approximate a hierarchical Kendall copula especially when the dependence is weak and if the vine copula is well chosen (note that only the results of the best fitting among the 15 considered vine copulas are shown in Figures \ref{fig:misspec} and \ref{fig:misspec2}).
The selection of vine copulas in higher dimensions is however a major problem (see \shortciteN{CzadoBrechmannGruber2013}) and the models typically do not stay as parsimonious and as easily interpretable as hierarchical Kendall copulas.
Finally, the results show that also elliptical and hierarchical Archimedean copulas can sometimes provide good approximations to hierarchical Kendall copulas in this four-dimensional setting.
It is, however, to be expected that the differences among the models become larger in higher dimensions.

%---------------------------------------------------------%

\section{Application}\label{sect:applic}

Finance is a major field, where copulas are used for dependence modeling (see, e.g., \shortciteN{CherubiniLucianoVecchiato2004}).
Often financial data exhibits some kind of clustering structure such as industry sectors and national stock markets.
For such data, hierarchical Kendall copulas are very suitable.
To investigate the usefulness of this newly proposed class of dependence models and to illustrate the presented inference techniques, the most important German stock market index DAX is analyzed.

The DAX is composed of 30 major German stocks.
For these we identified ten industry sectors: financials (Allianz, Commerzbank, Deutsche Bank, Deutsche B\"orse, Munich Re), chemicals (BASF, Bayer, K+S, Linde), healthcare (Fresenius, Fresenius Medical Care, Merck), automobile (BMW, Daimler, Volkswagen), industrials (MAN, Siemens, ThyssenKrupp), retail and consumer goods (Adidas, Beiersdorf, Henkel, Metro), IT and communications (Deutsche Telekom, Infineon, SAP), utilities (E.ON, RWE), transportation and logistics (Deutsche Post, Lufthansa), and building materials (HeidelbergCement).

\begin{figure}[t]
\centering
\includegraphics[width=.975\linewidth]{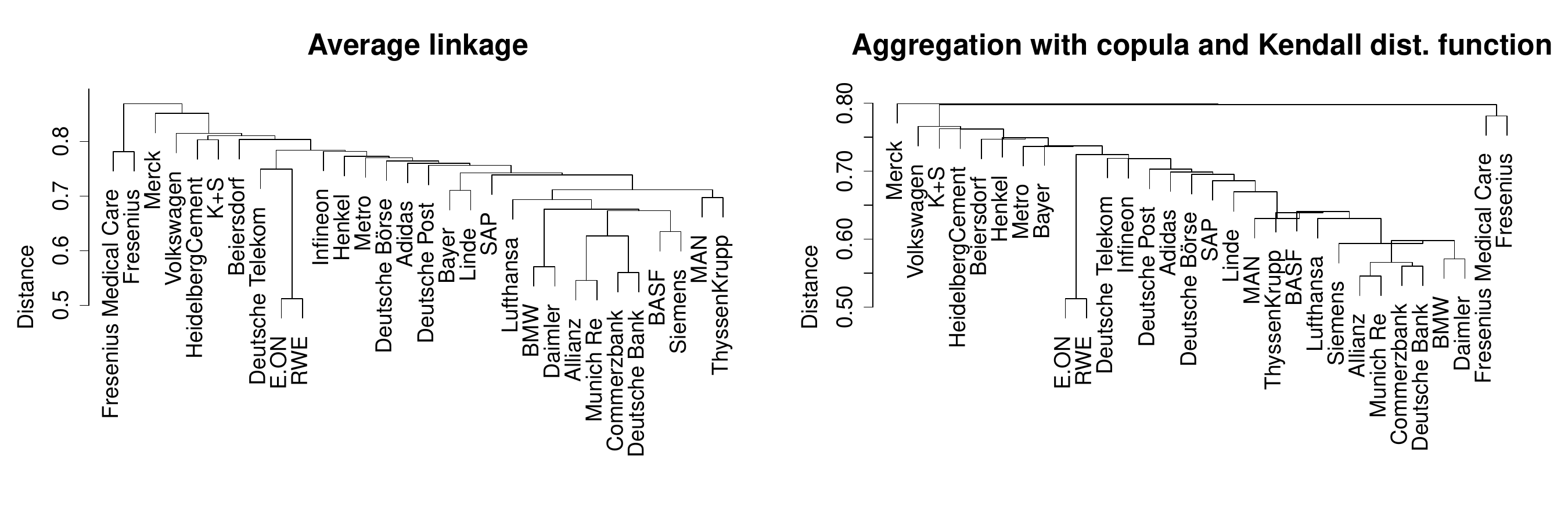}
\caption{Dendograms of the DAX constituents according to average linkage (left panel) and aggregation using the empirical Kendall distribution function of clusters (right panel).}
\label{fig:cluster}
\end{figure}

For all 30 stocks, more than six years of log returns (January 2005 to July 2011) are considered, where the time series are split into a training set of 1158 observations and a testing set of 500 observations for out-of-sample validation of our models.
As it is common for copula modeling in finance, we preliminarily fit time series models to the marginal time series and then work with standardized residuals which are transformed to marginally uniform data by the probability integral transform (inference functions for margins method; see Section \ref{sect:est}).
In particular, marginal GARCH(1,1)-models with Student's t innovations are chosen, which have been validated with appropriate tests.
Using this data, we illustrate the cluster selection procedures described in Section \ref{sect:modselect} and perform hierarchical clustering with the metric \eqref{eq:metric} and average linkage as well as aggregation using the empirical Kendall distribution function of clusters.
The resulting dendograms are shown in Figure \ref{fig:cluster}.
For instance, the utility and the healthcare sector can easily be identified.
This is not the case for the chemical and the IT companies.

Table \ref{tab:daxtautdep} shows the mean pairwise empirical Kendall's $\tau$ and the estimated degrees of freedom of a multivariate Student's t copula for each cluster.
Evidently, within-sector dependence is variable, since some clusters are more homogeneous than others.
Also strong tail dependence, as indicated by small degrees of freedom, cannot be found in all clusters.

\begin{table}[t]
\small
\begin{center}
\begin{tabular}{lrrrrrrrrr}
Sectors & Fin. & Chem. & Healthc. & Auto. & Ind. & Retail & IT & Util. & Transp. \\ 
\hline
\hline
Size & 5 & 4 & 3 & 3 & 3 & 4 & 3 & 2 & 2 \\
Mean pairw. Kendall's $\tau$ & 0.41 & 0.33 & 0.21 & 0.39 & 0.38 & 0.26 & 0.28 & 0.56 & 0.29 \\ 
Estimated deg. of freedom & 8.80 & 10.70 & 22.96 & 12.65 & 8.13 & 10.07 & 8.74 & 4.63 & 7.22 \\
\hline
\end{tabular}
\caption{Sector size, mean pairwise empirical Kendall's $\tau$ and estimated degrees of freedom of a Student's t copula for each cluster.}
\label{tab:daxtautdep}
\end{center}
\end{table}

\begin{table}[t]
\small
\centering
\begin{tabular}{lcccccc}
Model & \# Par. & Seq. Est. Log Lik. & Joint MLE Log Lik. & AIC & BIC\\
\hline
\hline
(Clayton, Student's t) & 55 & 6656.50 & 6677.73 & -13245.46 & -12967.47 \\ 
(Gumbel, Student's t) & 55 & 6989.32 & 6992.29 & -13874.57 & -13596.58 \\ 
(Frank, Student's t) & 55 & 7185.70 & 7190.29 & -14270.58 & \textbf{-13992.58} \\ 
\hline
(Clayton, Clayton) & 10 & 5452.34 & 5471.09 & -10922.17 & -10871.63 \\ 
(Gumbel, Gumbel) & 10 & 5860.74 & 5862.93 & -11705.85 & -11655.31 \\ 
(Frank, Frank) & 10 & 6003.60 & 6005.56 & -11991.12 & -11940.58 \\
\hline
Multivar. Gaussian & 435 & - & 8487.71 & \textbf{-16105.41} & -13906.73\\
Multivar. Student's t & 436 & - & 8906.14 & \textbf{-16940.28} & \textbf{-14736.54}\\
\hline
Regular vine & 509 & - & 9512.29 & \textbf{-18006.58} & \textbf{-15433.87}\\
\hline
\end{tabular}
\caption{Estimation results based on the training set. Notation of models: (cluster copulas, nesting copula).  AIC and BIC values are based on the joint MLE.}
\label{tab:daxresults}
\end{table}

We then fitted different hierarchical Kendall copulas to the training data set.
Results (log likelihood, AIC, BIC) are reported in Table \ref{tab:daxresults}.
As cluster copulas, we considered three different Archimedean copulas to account for different dependence structures as typically observed in financial data: Clayton with lower tail dependence, Gumbel with upper tail dependence and Frank with no tail dependence.
For the nesting copulas, we also investigated Gaussian and Student's t copulas, where the fits of the Gaussian copula to the aggregated pseudo observations of the different sectors were always inferior to that of the Student's t copula and are therefore not displayed here.
The different specified models are shown in the first column of Table \ref{tab:daxresults}.
The finite sample behavior of the estimates is investigated in Section \ref{sect:estsimstudy}.

For comparison, we also fitted classical multivariate Gaussian and Student's t copulas as well as a regular vine pair copula construction using the selection and estimation algorithm by \shortciteN{DissmannBrechmannCzadoKurowicka2013}, which is implemented in the R package \texttt{VineCopula} by \shortciteN{VineCopula}.
Pair copulas are selected from the following list: Gaussian, Student's t, Clayton, Gumbel
and Frank as well as rotations by $90$, $180$ and $270$ degrees of the tail-asymmetric Clayton and Gumbel copulas.
A 30-dimensional hierarchical Archimedean copula could however not be fitted due to the dependence restrictions of hierarchical Archimedean copulas (see Section \ref{sect:hierarch}): while there is moderate dependence within some clusters (see Table \ref{tab:daxtautdep}), there still is considerable and heterogeneous dependence among clusters (pairwise Kendall's $\tau$ values of aggregated data are ranging between 0.15 and 0.51), which cannot be modeled using a hierarchical Archimedean copula.

The hierarchical Kendall copulas therefore benefit from not having this restriction on within- and between-cluster dependence.
Moreover, ten-dimensional Student's t nesting copulas appear more reasonable than Archimedean nesting copulas (with only one parameter) due to the varying pairwise between-cluster dependence.
With respect to cluster copulas, the tail-symmetric Frank copula is, according to the AIC, superior to tail-asymmetric ones (Clayton, Gumbel).
For the two hierarchical Kendall copulas with Frank cluster copulas, parameter estimates and their standard errors according to the stationary bootstrap by \citeN{PolitisRomano1994} with an average block length of 20 observations can be found in Table \ref{tab:stderr}, showing that there is significant within- and between-sector dependence.
In particular, the within-sector dependence is quite variable and therefore can also not be appropriately fitted by an exchangeable 30-dimensional Archimedean copula.
For instance, the maximum likelihood parameter estimate of a 30-dimensional Frank copula is 2.096, of which most cluster copula parameters are significantly different.

\begin{table}[t]
\small
\begin{center}
\begin{tabular}{lllll}
& \multicolumn{2}{c}{(Frank, Student's t)} & \multicolumn{2}{c}{(Frank, Frank)}\\
& Estimate (Kendall's $\tau$) & Std. Error & Estimate (Kendall's $\tau$) & Std. Error \\ 
\hline
\hline
Financials  & 3.816 (0.374) & 0.210 & 3.946 (0.384) & 0.210 \\ 
Chemicals & 2.915 (0.300) & 0.197 & 3.047 (0.311) & 0.199 \\ 
Healthcare & 1.810 (0.195) & 0.191 & 2.003 (0.214) & 0.182 \\ 
Automobile & 3.775 (0.371) & 0.237 & 3.897 (0.380) & 0.238 \\ 
Industrials & 3.634 (0.360) & 0.303 & 3.880 (0.379) & 0.304 \\ 
Retail & 2.189 (0.232) & 0.168 & 2.394 (0.252) & 0.167 \\ 
IT and comm. & 2.401 (0.253) & 0.218 & 2.664 (0.277) & 0.215 \\ 
Utilities & 6.829 (0.555) & 0.544 & 6.912 (0.558) & 0.543 \\ 
Transportation & 2.669 (0.278) & 0.325 & 2.981 (0.306) & 0.329 \\ 
\hline
Between-sector & 0.248--0.704 (0.160--0.497) & 0.020--0.043 & 2.979 (0.305) & 0.165 \\ 
Degrees of freedom & 19.597 & 1.559 & - & - \\ 
\hline
\end{tabular}
\caption{Parameter estimates and their estimated standard errors of the hierarchical Kendall copulas with Frank cluster copulas (based on the training set). Kendall's $\tau$ values corresponding to the parameter estimates are given in brackets. For the entries of the Student's t correlation matrix ranges are reported.}
\label{tab:stderr}
\end{center}
\end{table}

In comparison to standard multivariate Gaussian and Student's t copulas, hierarchical Kendall copulas perform quite well, in particular when taking into account the enormous number of parameters of these models.
The number of parameters of elliptical copulas could be reduced significantly by using clustered correlation matrices.
Nevertheless, these have to fitted carefully in order to satisfy positive definiteness constraints.
Overall the regular vine copula provides the best fit, since it constitutes the most flexible model.
It is however even less parsimonious than the elliptical copulas and is not straightforward to interpret, especially not in terms of sectoral dependence.
Given that the highly parameterized multivariate elliptical copulas and regular vine pair copula constructions can be regarded as the current state-of-the-art models for financial return data, we focus on these models and investigate if the more parsimonious hierarchical Kendall copulas are competitive with them.
The good in-sample results obtained so far are in line with the misspecification study in Section \ref{sect:misspec}, where hierarchical Kendall copulas with elliptical nesting and Frank or Gumbel cluster copulas are shown to be reasonably close to multivariate elliptical models and, to some extent, to pair copula constructions.

\subsection{Value-at-Risk forecasting}

In finance, interest is however not so much in a good in-sample fit but rather in out-of-sample validation.
A typical exercise for this is Value-at-Risk (VaR) forecasting.
If the distribution of returns is continuous, the $(1-\alpha)$-VaR is the $\alpha$-quantile of the distribution.
For risk management, this value needs to be predicted on a daily basis, which we do for the testing set of 500 days using moving windows of length 1158.

Forecasts are typically evaluated in terms of exceedances, that is, the event that the predicted VaR is exceeded by the observed return.
For 500 forecasts, on average $500\times \alpha$ exceedances are expected.
Whether the number of exceedances (``unconditional coverage'') and their occurrences (should be independent; both properties: ``conditional coverage'') are appropriate can be evaluated using a range of tests, so-called backtests, that have been proposed in the literature: the proportion of failures test of unconditional coverage by \citeN{Kupiec1995} (UC), the Markov test of independence by \citeN{Christoffersen1998} (IND1), the joint test of conditional coverage by \citeN{Christoffersen1998} (CC1), the mixed Kupiec test of conditional coverage by \citeN{Haas2001} (CC2), the Weibull test of independence by \citeN{ChristoffersenPelletier2004} (IND2), and the duration-based GMM test of conditional coverage by \shortciteN{CandelonColletazHurlinTokpavi2011} (CC3 and CC4 with orders 2 and 5, respectively).

\begin{table}[t]
\small
\begin{center}
\begin{tabular}{lccrrrrrrr}
Model & Level & \# Exceed. & UC & IND1 & IND2 & CC1 & CC2 & CC3 & CC4 \\ 
\hline
\hline
Independence & 99\% & 103 & \textit{0.00} & \textit{0.04} & \textit{0.00} & \textit{0.00} & \textit{0.00} & \textit{0.00} & \textit{0.00} \\ 
copula & 95\% & 135 & \textit{0.00} & 0.15 & \textit{0.03} & \textit{0.00} & \textit{0.00} & \textit{0.00} & \textit{0.00} \\ 
 & 90\% & 157 & \textit{0.00} & 0.17 & \textit{0.03} & \textit{0.00} & \textit{0.00} & \textit{0.00} & \textit{0.00} \\ 
\hline
Gaussian & 99\% & 5 & 1.00 & 0.75 & 0.79 & 0.95 & 0.35 & 0.93 & 1.00 \\ 
copula & 95\% & 26 & 0.84 & 0.09 & 0.74 & 0.23 & 0.35 & 0.98 & 1.00 \\ 
 & 90\% & 52 & 0.77 & 0.46 & 0.87 & 0.73 & \textit{0.03} & 0.29 & 0.73 \\ 
\hline  
Student's t & 99\% & 4 & 0.64 & 0.80 & 0.21 & 0.87 & 0.93 & 0.75 & 0.95 \\ 
copula & 95\% & 24 & 0.84 & 0.12 & 0.73 & 0.29 & 0.39 & 0.89 & 1.00 \\ 
 & 90\% & 53 & 0.66 & 0.53 & 0.85 & 0.74 & 0.05 & 0.31 & 0.73 \\ 
\hline
Regular vine & 99\% & 5 & 1.00 & 0.75 & 0.81 & 0.95 & 0.35 & 0.93 & 1.00 \\ 
pair copula & 95\% & 27 & 0.69 & 0.08 & 0.70 & 0.20 & 0.35 & 0.95 & 1.00 \\ 
construction & 90\% & 53 & 0.66 & 0.53 & 0.87 & 0.74 & \textit{0.04} & 0.24 & 0.66 \\ 
\hline   
Hierarchical & 99\% & 2 & 0.13 & 0.90 & 0.46 & 0.31 & 0.35 & 0.43 & 0.82 \\ 
Kendall copula & 95\% & 31 & 0.23 & \textit{0.04} & 0.76 & 0.06 & 0.21 & 0.55 & 0.92 \\ 
(Frank, Student's t) & 90\% & 57 & 0.31 & 0.52 & 0.81 & 0.48 & \textit{0.03} & 0.22 & 0.64 \\    
\hline
\end{tabular}
\caption{$P$-values of VaR backtests for hypotheses of independence and (un)conditional coverage. We expect 5/25/50 exceedances at the 99\%/95\%/90\%-level, respectively.}
\label{tab:daxVaR}
\end{center}
\end{table}

Here, the Value-at-Risk of an equally weighted portfolio of the 30 DAX stocks is forecasted.
Backtesting results of the following five different models can be found in Table \ref{tab:daxVaR}:
\begin{itemize}

\item Independence copula (for comparison);

\item Gaussian and Student's t copulas;

\item Regular vine pair copula construction;

\item Hierarchical Kendall copula with Frank cluster copulas and with Student's t nesting copula (best fit among the considered hierarchical Kendall copulas).

\end{itemize}
In summary, none of the hypotheses of independence and (un)conditional coverage can  consistently be rejected for any of the VaR levels and for any of the models---except for the multivariate independence copula, as expected.
The weak lack of conditional coverage at the 90\% level, as detected by the mixed Kupiec test of \citeN{Haas2001}, is not supported by the other tests.

This shows that hierarchical Kendall copulas are as good as the common Gaussian and Student's t copulas and also as the more flexible pair copula constructions when it comes to out-of-sample validation.
In particular, the hierarchical Kendall copula with Frank cluster copulas, which we consider here, is very parsimonious and allows for closed-form calculations and very efficient simulations due to its Archimedean clusters.
In contrast to elliptical and vine copulas, it is also directly interpretable in terms of within- and between-sector dependence.

%---------------------------------------------------------%

\section{Discussion}\label{sect:concl}

In this paper we introduce and discuss the new class of hierarchical Kendall copulas.
By grouping variables at different hierarchical levels, it provides an appealing construction principle for high-dimensional dependence models.
It is shown that the important special cases of independence as well as of comonotonicity belong to this model class.
For Archimedean cluster copulas, a stochastic representation is given and differences to hierarchical Archimedean copulas are investigated.
Most importantly, the density of hierarchical Kendall copulas is derived.

Thereafter, statistical inference techniques for hierarchical Kendall copulas are developed.
In particular, simulation algorithms are provided, with focus on Archimedean clusters, for which convenient closed-form expressions are derived.
The availability of the density of hierarchical Kendall copulas renders feasible estimation using maximum likelihood techniques.
Finally, we show that a model with Archimedean cluster copulas and Student's t nesting copula performs very well in a substantial financial application.

Nevertheless, it is a open research question to derive sampling and estimation methods for the case that copulas and Kendall distribution functions are not available in closed form.
In particular, the popular class of elliptical copulas is of interest here.

%---------------------------------------------------------%

\section*{Acknoweledgement}

The author gratefully acknowledges the helpful comments of the referees, which further improved the manuscript.
He also likes to thank his supervisor Claudia Czado as well as Harry Joe, Johanna Ne\v{s}lehov\'a, Matthias Scherer and Michael Smith for instructive remarks and discussions.
Furthermore, he acknowledges the support of the TUM Graduate School's International School of Applied Mathematics as well as of Allianz Deutschland AG.
The numerical computations were performed on a Linux cluster supported by DFG grant INST 95/919-1 FUGG. 

%---------------------------------------------------------%

\markboth{Bibliography}{Bibliography}
\addcontentsline{toc}{section}{Bibliography}
\bibliography{hier_cop-bib}

\end{document}